# The features of a quantum description of radiation in an optically dense medium

M. D. Tokman, M. A. Erukhimova, and V. V. Vdovin

Institute of Applied Physics of the Russian Academy of Sciences
46 Ul'yanov Street, 603950 Nizhny Novgorod, Russia
e-mail: tokman@appl.sci-nnov.ru

**Corresponding author:**
M.A. Erukhimova, e-mail: eruhmary@appl.sci-nnov.ru, ph. +78314164754,
Institute of Applied Physics of the Russian Academy of Sciences
46 Ul'yanov Street, 603950 Nizhny Novgorod, Russia

**Abstract:**
This paper is devoted to the theory of quantum electromagnetic field in an optically dense medium. Self-consistent equations describing interaction between a quantum field and a quantum dielectric medium are obtained from the first principles, i.e., outside a phenomenological description. Using these equations, we found a transformation (of the Bogoliubov transformation type) that converts the operators of the "vacuum" field into operators of collective perturbations of the field and an ensemble of atoms, that is, photons in the medium. Transformation parameter is the refractive index of the wave mode considered. It is shown that besides the energy of the collective electromagnetic field, the energy of photons in the medium includes the energy of the internal degrees of freedom of the substance and the energy of near-field dipole interaction between atoms in the polarized medium. The concept of negative energy photons is introduced on the basis of self-consistent equations.

**Keywords:** Optically dense medium, quantum field, Bogoliubov transformation, photons in the medium

I. INTRODUCTION

Quantum theory of light in the medium is one of the most important areas of applied physics [1-9]. Currently, research in this field is largely associated with the use of the properties of the wave functions of "multiparticle" systems for the implementation of some algorithms for computing, secure communication channels, and precision measurements [9-14].

Quantum description of the electromagnetic field in the medium is usually based on the application of a standard quantization procedure to the phenomenological equations of classical electrodynamics of a continuous medium [2, 15-21]. The theory is phenomenological in that specified dielectric permittivity and/or magnetic permeability are assigned to the medium [22,23]. This procedure is obviously not always adequate for the problem of self-consistent description of the field and the medium. Self-consistent analysis of interaction between the quantum field and the medium is usually performed within the framework of the perturbation method, which assumes that the wave number of the radiation differs little from the "vacuum" value $\omega/c$ (here, $c$ is the speed of light in va-cuum). For example, for the medium with polarizability $\chi$ and refractive index $(ck/\omega)^2 \equiv n^2 = 1 + 4\pi\chi$ the result obtained within this approach corresponds to the approximation of small optical density, $n \approx 1 + 2\pi\chi$. Despite the obvious constraints, this approach can be very effective. The point is that the approximation of small optical depth, i.e., the assumption that $|n-1| \ll 1$ in a fairly narrow frequency band, does not contradict, in principle, the condition of strong frequency dispersion, $\omega|n'_\omega| \gg 1$. As an example, we mention the effect of electromagnetically induced transparency (EIT) [24-27]: which suggests that within the "transparency window" there is a frequency range in which a small group velocity is combined with an almost "vacuum" phase velocity. However, in general (including some of the EIT modes, see [26, 27]), the constraint $|n-1| \ll 1$ is awkward. To our knowledge, the analysis which is free of both the constraints of the phenomenological approach and the small optical density approximation, was carried out only in terms of a two-level model [5, 28-30].

This paper is devoted to the development of the theory of quantum field in a medium with an arbitrary optical density and an arbitrary energy-level structure. Selecting as the initial model an ensemble of atoms interacting through a collective field, we came to fairly universal operator equations of quantum electrodynamics of a dielectric medium without spatial dispersion. Using these equations, it was found that the exact dispersion relation $\omega = \omega(k)$ for photons in the medium corresponds to the quanta of collective excitations of the field and the medium, and the energy $\hbar\omega(k)$ of a quantum includes the energy of the macroscopic field, the energy of the internal degrees of freedom, and the energy of the near-field dipole interaction in the polarized medium. It is shown that the operators of creation and annihilation the photon in the medium are related with the "vacuum" operators by the Bogoliubov transformation [31, 32] with the refractive index of the medium as the transformation parameter. The concept of negative energy quanta of the wave disturbances was also introduced on the basis of the analysis of self-consistent operator equations for the field in the medium.

The paper is organized as follows. In Sections **II** and **III** we formulate the operator equations, that describe the interaction between quantum field and multi-level medium, taking into account the local-field effect that is typical of the optically dense media [5,6,22,33]. In Section **IV** we define the linear response of the medium to the quantized field. In Section **V** we pass to self-consistent quantum equations for the field and a continuous transparent medium in the secondary quantization representation. In Sections **VI** and **VII** we study the properties of the quanta of collective excitations of the field and the medium, in particular, their representation by the Bogoliubov transformation of the "vacuum" photon operators. Sections **VIII** and **IX** are devoted to the concept of negative energy photons and generalization of obtained relations to the case of media driven by the classical pump. In Appendix **A** we came to fairly universal operator equations of quantum electrodynamics of a dielectric medium without spatial dispersion, selecting as the initial model an ensemble of atoms interacting through a collective field; Appendix **B** is devoted to the specificity of initial conditions formulation for the operators describing quantum medium.

## II. GENERAL EQUATIONS FOR QUANTUM FIELD IN A TRANSPARENT CONTINOUS MEDIUM.

Let the considered medium be characterized by the energy spectrum $W_n$ and the set of matrix elements $f_{mn}$ for any physical quantity $f$ of interest to us. The subscripts $n$, $m$ are supposed to be discrete although this is not essential. For a continuous medium without spatial dispersion, one could directly introduce the statistical matrix $\rho_{mn}(r,t)$ that characterizes the internal degrees of freedom and depends on the coordinate $r$ of the point in space as on the parameter. Then the spatial-temporal dependence of the averaged value $f(r,t)$ is defined by relation $f(r,t) = \sum_{n,m} f_{nm} \rho_{mn}(r,t)$ [34-36]. Considering the interaction of medium with quantum field it is convenient to introduce the Heisenberg operator for the distributed quantity $\hat{f}(r,t)$ that is defined via corresponding Heisenberg density operator: $\hat{f}(r,t) = \sum_{n,m} f_{nm} \hat{\rho}_{mn}(r,t)$. The following commutation relation is fulfilled for the operator $\hat{\rho}_{mn}(r,t)$ [37]:

$$[\hat{\rho}_{qp}(r,t), \hat{\rho}_{mn}(r',t)] = \delta(r - r')(\delta_{qn} \hat{\rho}_{mp}(r,t) - \delta_{mp} \hat{\rho}_{qn}(r,t)) \quad (1)$$

The operators $\hat{f}(r,t)$ and $\hat{\rho}_{mn}(r,t)$ are determined as corresponding operators for separate atoms (or the other elementary medium "cells") averaged over physically small volume; the consistent averaging procedure is presented in Appendix **A** (see also [25,26,37]).

Further we use the relations which are quite clear from the physical point of view, their proof is given in Appendix **A**:

$$\hat{H} = \int_V \left( \frac{\hat{E}^2 + \hat{B}^2}{8\pi} - \frac{\alpha}{2} \hat{P}^2 + \hat{w} \right) d^3 r \quad (2a)$$

It is the quantum operator of the energy of the system that includes field and medium, $V$ is the quantization volume, $\hat{E}(r,t)$ and $\hat{B}(r,t)$ are the operators of electric and magnetic field;

$$\hat{w}(r,t) = \sum_n W_n \hat{\rho}_{nn}(r,t) \quad (2b)$$

is the operator of the internal energy of the medium;

$$\hat{P}(r,t) = \sum_{n,m} d_{nm} \hat{\rho}_{mn}(r,t) \quad (2c)$$

is the operator of polarization of the unit volume. The presence of a term proportional to the parameter $\alpha$ in Eq. (2a) is related to allowance of the energy of the near dipole interaction in the polarized medium (the most widespread Lorenz-Lorentz model corresponds to the value $\alpha = 4\pi/3$) [5,6,22,33,38].

As usual, the operator of vector potential $\hat{A}$ can be taken as the operator of canonical coordinate of the field, having in mind the Coulomb gauge $\nabla \hat{A} = 0$. In frame of electric-dipole field-medium interaction the operator of momentum $\hat{F}$ canonically conjugated to the chosen coordinate is proportional to the operator of electric displacement $\hat{D}$:

$$\hat{F} = -\frac{\hat{D}}{4\pi c}, \quad (3a)$$

$$\hat{D} = \hat{E} + 4\pi \hat{P}. \quad (3b)$$

The proof of relation (3a) is given in Appenvix **A** (see also [2,5,15]). After transformation to the canonical variables the Hamiltonian (2a) takes the form:

$$\hat{H} = \int_V \left( \frac{\hat{D}^2 + (\nabla \times \hat{A})^2}{8\pi} + \hat{w} - \hat{D}\hat{P} + \left(2\pi - \frac{\alpha}{2}\right)\hat{P}^2 \right) d^3r; \quad (4)$$

Operators of the canonically conjugate variables $\hat{F}$ and $\hat{A}$ will be specified by a standard commutation relation for the canonically conjugate "momentum – coordinate" pair [1,4,5]:

$$[\hat{F}_\alpha(\mathbf{r}',t), \hat{A}_\beta(\mathbf{r},t)] = -i\hbar \delta_{\alpha\beta} \delta(\mathbf{r} - \mathbf{r}'), \quad (5a)$$

where $\alpha, \beta = x, y, z$ are the indices of the Cartesian coordinates of the vectors. Since the commutation relations for the Heisenberg operators are preserved and equal to the corresponding relations for the Schrodinger operators [4,39], the field and medium operators commute with each other:

$$[\hat{F}_\alpha(\mathbf{r}',t), \hat{\rho}_{mn}(\mathbf{r},t)] = [\hat{A}_\alpha(\mathbf{r}',t), \hat{\rho}_{nm}(\mathbf{r},t)] = 0. \quad (5b)$$

Taking into account the following identity, derived in Appendix **A**:

$$\left[ \int_V (\hat{D}\hat{P}) d^3r, \hat{A} \right] = \left[ \int_V (\hat{D}\hat{P}_\perp) d^3r, \hat{A} \right], \quad (5c)$$

it can be verified that Hamiltonian (4) corresponds to the result given in [5,28,29] for the particular case of a two-level system. In (5c) the representation of the polarization vector as the sum of the vortex and potential components is used: $\hat{P} = \hat{P}_\perp + \hat{P}_\parallel$, where $\nabla \times \hat{P}_\parallel = \nabla \hat{P}_\perp = 0$.

We use the Heisenberg equations for the operators of momentum and coordinate of the field. Thus, from the equation

$$\dot{\hat{F}} = \frac{i}{\hbar}[\hat{H}, \hat{F}],$$

using expressions (4) and (5a), we obtain

$$\dot{\hat{\boldsymbol{D}}} = c\nabla \times \nabla \times \hat{\boldsymbol{A}}.$$

(from this equality we find, in particular, the condition $\nabla \hat{\boldsymbol{D}} = 0$ which is required to prove identity (5c)). From the equation

$$\dot{\hat{\boldsymbol{A}}} = \frac{i}{\hbar}[\hat{H}, \hat{\boldsymbol{A}}]$$

in view of relation (5c) we have

$$\dot{\hat{\boldsymbol{A}}} = -c(\hat{\boldsymbol{D}} - 4\pi\hat{\boldsymbol{P}}_\perp).$$

Taking into account the Coulomb calibration, we arrive at an operator analogue of wave equation:

$$\ddot{\hat{\boldsymbol{D}}} + c\nabla^2 \dot{\hat{\boldsymbol{A}}} = 0. \tag{6a}$$

For the electric field we obtain the following identities:

$$\hat{\boldsymbol{E}} = (\hat{\boldsymbol{D}} - 4\pi\hat{\boldsymbol{P}}) = -\frac{\dot{\hat{\boldsymbol{A}}}}{c} - 4\pi\hat{\boldsymbol{P}}_\| = -\frac{\dot{\hat{\boldsymbol{A}}}}{c} - \nabla\hat{\varphi}; \tag{6b}$$

the latter relation corresponds to the operator counterpart of the Poisson equation:

$$\nabla\hat{\varphi} = 4\pi\hat{\boldsymbol{P}}_\|. \tag{6c}$$

Equations (6a), (6b), and (6c) conform to the general rule of identity of classical evolution equation and the Heisenberg equations for the operators.

The Heisenberg equation for the density operator of a distributed system

$$\dot{\hat{\rho}}_{mn} = \frac{i}{\hbar}[\hat{H}, \hat{\rho}_{mn}]$$

with allowance for Eqs. (4), (2b,c), and (1) leads to the evolution equation:

$$\dot{\hat{\rho}}_{mn} = -\frac{i}{\hbar}\sum_\nu (\hat{h}_{m\nu}\hat{\rho}_{\nu n} - \hat{\rho}_{m\nu}\hat{h}_{\nu n}), \tag{7a}$$

where

$$\hat{h}_{m\nu}(\boldsymbol{r},t) = W_m\delta_{m\nu} - \hat{\boldsymbol{E}}^a(\boldsymbol{r},t)\boldsymbol{d}_{m\nu}, \tag{7b}$$

$$\hat{\boldsymbol{E}}^a = \hat{\boldsymbol{E}} + \alpha\hat{\boldsymbol{P}} = \hat{\boldsymbol{D}} - (4\pi - \alpha)\hat{\boldsymbol{P}}, \tag{7c}$$

$\hat{\boldsymbol{E}}^a(\boldsymbol{r},t)$ is the local-field operator [5,6,22,33].

Dynamic operator equations (6a), (6b), (6c) and (7a), (7b), (7c) with connection (2c) describe the Hamiltonian system "quantum field + dielectric quantum medium". Sometimes it is needed to generalize the obtained equations to the case of open systems, which are characterized by dissipative effects. This can be done by adding in an additive way the relaxation and Langevin noise operators to the dynamics equation (7a) of an atomic system (see, e.g., [16,24-26,37,40,41,43]):

$$\dot{\hat{\rho}}_{mn} = -\frac{i}{\hbar}\sum_{\nu}\left(\hat{h}_{m\nu}\hat{\rho}_{\nu n} - \hat{\rho}_{m\nu}\hat{h}_{\nu n}\right) + \hat{R}_{mn} + \hat{L}_{mn}. \qquad (7d)$$

Here, $\hat{R}_{mn}$ is the relaxation operator, $\hat{L}_{mn}$ is the Langevin noise operator describing fluctuations in the atomic system. Among the next possible generalizations of Eqs. (7a) and (7d), we mention allowance for the distribution of atoms over orientations of their symmetry axes, transition from a discrete set of numbers $n$ to a continuous spectrum, and rejection of the electric dipole approximation. In general, however, Eqs. (6a), (6b), (6c), (7a), (7b), (7c) and (2c) make up a fairly universal model, which permits one to discuss a number of fundamental aspects of quantum electrodynamics of a continuous medium.

## III. THE INITIAL CONDITIONS FOR THE OPERATOR EQUATIONS

In frame of Heisenberg description the operators act on the fixed state vector $|\Psi\rangle$ and each Heisenberg operator is the function of time and initial conditions. The corresponding initial values are the Schrödinger (constant) operators of the system variables. It is convenient to assume that the state vector for the "atom+medium" system is defined by simple relation $|\Psi\rangle = |\Psi_F\rangle|\Psi_A\rangle$, where $|\Psi_F\rangle$ is field state and $|\Psi_A\rangle$ is the medium sate which are independent (all possible complications can be taken into account by the corresponding recalculation of initial values for operators, see Appendix **B** for details). For the considered system (6a), (6b), (6c) and (7a), (7b), (7c) with connection (2c) interaction between the field and the medium leads naturally to the entanglement of these subsystems. The dynamics of such entanglement under Heisenberg approach appears as the time-dependent functional dependence between operators corresponding to different subsystems.

For different specific problems it is useful to reduce the complete system to one of subsystems. It is realized by the averaging over variables of the other subsystem. In the problems of quantum optics, it is accepted to use field operators,[1] averaged over the initial state of the medium. In such approach it is convenient to specify the initial density operator $\hat{\rho}^0_{mn}(\mathbf{r})$ not directly, but via mean values at the initial time that are necessary for the problem. For example, $\langle\Psi_A|\hat{\rho}^0_{mm}(\mathbf{r})|\Psi_A\rangle = N_m(\mathbf{r})$ is the initial spatial distribution of populations, $\langle\Psi_A|\hat{\rho}^0_{m\neq n}(\mathbf{r})|\Psi_A\rangle = \sigma_{mn}(\mathbf{r})$ is the initial spatial distribution of quantum coherences, $\langle\Psi_A|\hat{\rho}^0_{mn}(\mathbf{r})\hat{\rho}^0_{pq}(\mathbf{r}')|\Psi_A\rangle$ is the initial value of the correlator, etc. (Here $\hat{\rho}^0_{mn}(\mathbf{r}) = \hat{\rho}_{mn}(t=0,\mathbf{r})$).

## IV. LINEAR SOLUTION OF THE MEDIUM DYNAMICS EQUATIONS

---

[1] In [5], the opposite case, i.e., the transition to operators acting only on the state of the medium, is also considered.

Consider a linear response of the medium to the action of the quantum field. From Eqs. (7a), (7b), (7c), and (2c) within the framework of the linear approximation over the field $\hat{E}$ we obtain

$$\left.\begin{array}{l}\dot{\hat{\rho}}_{mn}+i\omega_{mn}\hat{\rho}_{mn}=\dfrac{i}{\hbar}\left(\left(\hat{\rho}_{nn}^{0}-\hat{\rho}_{mm}^{0}\right)\boldsymbol{d}_{mn}+\sum\limits_{\nu\ne m,n}\left(\boldsymbol{d}_{m\nu}\hat{\rho}_{\nu n}^{0}\,\mathrm{e}^{-i\omega_{\nu n}t}-\hat{\rho}_{m\nu}^{0}\,\mathrm{e}^{-i\omega_{m\nu}t}\boldsymbol{d}_{\nu n}\right)\right)\hat{\boldsymbol{E}}^{a}\\ \hat{\boldsymbol{E}}^{a}=\hat{\boldsymbol{E}}+\alpha\hat{\boldsymbol{P}},\quad \hat{\boldsymbol{P}}=\sum\limits_{m,n}\boldsymbol{d}_{nm}\hat{\rho}_{mn}\end{array}\right\},\quad(8)$$

where $\omega_{mn}=(W_m-W_n)/\hbar$. Assume that at the initial time the quantum coherence is absent. Next we consider forced oscillations of the medium in a monochromatic quantum field: $\hat{\boldsymbol{E}}=\hat{\boldsymbol{E}}_\omega(\boldsymbol{r})\mathrm{e}^{-i\omega t}+H.C.$ In the range of applicability of linear approximation over field $\hat{\boldsymbol{E}}$ it is convenient to make averaging over medium variables in equations for operators directly. Taking into account Eq. (B3) the diagonal elements of density operators $\hat{\rho}_{mm}$ will be substituted by constant populations (c-numbers) $N_m$, and the equations for off-diagonal elements $\hat{\rho}_{m\ne n}$ will depend on field operators and populations.[2]. As a result, from expressions (8) we obtain

$$\left.\begin{array}{l}\hat{\rho}_{\omega;mn}(\boldsymbol{r})=\dfrac{1}{\hbar}\sum\limits_{m,n}\dfrac{N_n(\boldsymbol{r})-N_m(\boldsymbol{r})}{\omega_{mn}-\omega}\boldsymbol{d}_{mn}\left(\hat{\boldsymbol{E}}_\omega(\boldsymbol{r})+\alpha\hat{\boldsymbol{P}}_\omega(\boldsymbol{r})\right)\\ \hat{\boldsymbol{P}}=\hat{\boldsymbol{P}}_\omega(\boldsymbol{r})\mathrm{e}^{-i\omega t}+H.C.,\quad \hat{\boldsymbol{P}}_\omega(\boldsymbol{r})=\sum\limits_{m,n}\boldsymbol{d}_{nm}\hat{\rho}_{\omega;mn}(\boldsymbol{r})\end{array}\right\}.\quad(9\mathrm{a})$$

Here and after we have in mind $\hat{\boldsymbol{P}},\hat{\boldsymbol{E}}\Rightarrow\langle\Psi_A|\hat{\boldsymbol{P}}|\Psi_A\rangle,\langle\Psi_A|\hat{\boldsymbol{E}}|\Psi_A\rangle$.

From Eqs.(9a) we obtain the following connection relations:

$$\left.\begin{array}{l}\hat{\boldsymbol{P}}=\vec{\chi}_\omega(\boldsymbol{r})\hat{\boldsymbol{E}}_\omega(\boldsymbol{r})\mathrm{e}^{-i\omega t}+H.C.\\ \vec{\chi}_\omega(\boldsymbol{r})=\left[\vec{1}-\alpha\vec{\chi}_\omega^{\,a}(\boldsymbol{r})\right]^{-1}\vec{\chi}_\omega^{\,a}(\boldsymbol{r})\end{array}\right\}.\quad(9\mathrm{b})$$

Here, $\vec{\chi}_\omega^{\,a}(\boldsymbol{r})$ is the tensor which we determine by its action on a certain vector $\boldsymbol{V}$:

$$\vec{\chi}_\omega^{\,a}(\boldsymbol{r})\boldsymbol{V}\equiv\dfrac{1}{\hbar}\sum\limits_{m,n}\dfrac{N_n(\boldsymbol{r})-N_m(\boldsymbol{r})}{\omega_{mn}-\omega}\boldsymbol{d}_{nm}(\boldsymbol{d}_{mn}\boldsymbol{V}),\quad(9\mathrm{c})$$

and $\vec{1}$ is the unit diagonal tensor. The tensor $\vec{\chi}_\omega(\boldsymbol{r})$ characterizes the polarizability of the medium in a monochromatic field.

Relaxation processes can be taken into account by replacing Eq.(7a) by Eq.(7d). Allowing for the property $\langle\Psi_A|\hat{L}_{mn}|\Psi_A\rangle=0$ of the Langevin operator [40], for the simplest relaxation operator with $\hat{R}_{m\ne n}=-\gamma_{mn}\hat{\rho}_{mn}$ we arrive at an expression differing from Eq.(9c) only by the replacement $\omega_{mn} \Rightarrow \omega_{mn} - i\gamma_{mn}$. The question about correctness of different forms of presentation of the relaxation operator was explored in [41,43,44-47].

---

[2] In [37,42] such approach was used in the system with classical drive field.

Exactly as in classical electrodynamics of continuous media (see, e.g., [2, 19, 23]), one can pass from the above relation between the monochromatic components of the field $\hat{\boldsymbol{E}}_\omega$ and the polarization $\hat{\boldsymbol{P}}_\omega$ to the more general relations

$$\hat{\boldsymbol{P}}(\boldsymbol{r},t) = \vec{\vec{\chi}}\hat{\boldsymbol{E}}(\boldsymbol{r},t) \equiv \int_0^\infty \vec{\vec{\chi}}(\boldsymbol{r},\tau)\hat{\boldsymbol{E}}(\boldsymbol{r},t-\tau)d\tau, \qquad (10a)$$

where

$$\int_0^\infty \vec{\vec{\chi}}(\boldsymbol{r},\tau)e^{i\omega\tau}d\tau = \vec{\vec{\chi}}_\omega(\boldsymbol{r}); \qquad (10b)$$

i.e., the tensor $\vec{\vec{\chi}}_\omega(\boldsymbol{r})$ defined by expressions (9b) and (9c) is a spectral image of the tensor integral operator[3] $\vec{\vec{\chi}}$ specified by the kernel $\vec{\vec{\chi}}(\boldsymbol{r},\tau)$.

We now apply the averaging procedure over the initial state $|\Psi_A\rangle$ of the atomic system to operator equations (6a) and (6b) for the field and then make use of relations (10a) and (10b). As a result, we obtain

$$\left.\begin{aligned} &\ddot{\hat{\boldsymbol{D}}} + c^2 \nabla \times \nabla \times \hat{\boldsymbol{E}} = 0 \\ &\hat{\boldsymbol{D}}(\boldsymbol{r},t) = \vec{\vec{\varepsilon}}\hat{\boldsymbol{E}}(\boldsymbol{r},t) \equiv \int_0^\infty \vec{\vec{\varepsilon}}(\boldsymbol{r},\tau)\hat{\boldsymbol{E}}(\boldsymbol{r},t-\tau)d\tau \end{aligned}\right\}, \qquad (10c)$$

where $\vec{\vec{\varepsilon}}$ is the tensor operator of dielectric permittivity;

$$\int_0^\infty \vec{\vec{\varepsilon}}(\boldsymbol{r},\tau)e^{i\omega\tau}d\tau = \vec{\vec{\varepsilon}}_\omega(\boldsymbol{r}) = \vec{1} + 4\pi\vec{\vec{\chi}}_\omega(\boldsymbol{r}). \qquad (10d)$$

Equations (10c) and (10d), which describe the field in the medium, coincide in form with the equations of the phenomenological theory of quantum field in a transparent medium [2,16-20]. However, under the approach we used here, the constitutive equations combined with the field equations follow from a closed quantum model, while the phenomenological theory defines the constitutive equations *a priori*.

V. REPRESENTING THE EQUATIONS FOR THE FIELD IN THE MEDIUM BY THE PHOTON CREATION AND ANNIHILATION OPERATORS

---

[3]To avoid a misunderstanding, we emphasize that the appearance of the operator $\vec{\vec{\chi}}$ is due to the specific formation of the electrodynamic response in a frequency-dispersive medium, rather than due to the quantum consideration [2,19,23].

We represent the Hamiltonian of the quantum field in the medium using the standard creation and annihilation operators of the Fock states [1,2,4,5,7,48]. Considering, as usual, periodic boundary conditions at the border of the quantization volume $V$, we obtain the following expressions for the operators of the field coordinate $\hat{A}$ and its canonical momentum $\hat{F} = -\hat{D}/4\pi c$:

$$\hat{A} = \sum_{k,\sigma} \left( A^0_{\sigma;k} e^{ikr} \hat{b}_{\sigma;k} + A^{0*}_{\sigma;k} e^{-ikr} \hat{b}^\dagger_{\sigma;k} \right), \tag{11a}$$

$$\hat{F} = -\frac{\hat{D}}{4\pi c} = -\frac{1}{4\pi c} \sum_{k,\sigma} \left( D^0_{\sigma;k} e^{ikr} \hat{b}_{\sigma;k} + D^{0*}_{\sigma;k} e^{-ikr} \hat{b}^\dagger_{\sigma;k} \right). \tag{11b}$$

Here, $A^0_{\sigma;k}$ and $D^0_{\sigma;k}$ are the normalization vector amplitudes, $k$ are the wave vectors determined by the boundary conditions, and the subscript $\sigma$ marks the polarization of the vector potential, so that[4]

$$A^0_{\sigma;k} A^{0*}_{\sigma';k} \propto \delta_{\sigma\sigma'},\ A^0_{\sigma;k} \perp k,\ A^0_{\sigma;-k} = A^{0*}_{\sigma;k}. \tag{11c}$$

The creation and annihilation operators of the corresponding Fock states $\hat{b}_{\sigma;k}$ and $\hat{b}^\dagger_{\sigma;k}$ satisfy the commutation relations for bosons:

$$[\hat{b}_{\sigma;k},\hat{b}_{\sigma';k'}] = [\hat{b}^\dagger_{\sigma;k},\hat{b}^\dagger_{\sigma';k'}] = 0,\quad [\hat{b}_{\sigma;k},\hat{b}^\dagger_{\sigma';k'}] = K_k \delta_{kk'} \delta_{\sigma\sigma'}. \tag{11d}$$

Relation between the normalization quantities $K_k$, $A^0_{\sigma;k}$, and $D^0_{\sigma;k}$ is determined, firstly, by commutation relation (5a), whence we obtain

$$D^{0*}_{\sigma;k} A^0_{\sigma';k} = -\frac{2i\pi c\hbar}{K_k V} \delta_{\sigma\sigma'}. \tag{12a}$$

Secondly, in the limiting case of vacuum ($\hat{P} = 0$) the Heisenberg equations of motion for the operators $\hat{b}_{\sigma;k}$ and $\hat{b}^\dagger_{\sigma;k}$ should correspond to the dispersion relation $\omega^2 = c^2 k^2$. As a result, we obtain

$$\left| D^0_{\sigma;k} \right|^2 + k^2 \left| A^0_{\sigma;k} \right|^2 = \frac{4\pi\hbar c|k|}{K_k V}. \tag{12b}$$

Besides, from the condition $\nabla\hat{D} = 0$ and relations (12a) it also follows that

$$D^0_{\sigma;k} \perp k,\ D^0_{\sigma;-k} = -D^{0*}_{\sigma;k}. \tag{12c}$$

Using the above expansions of the operators over spatial harmonics, we transform Hamiltonian (4) to the following form:

---

[4] Complex vectors $A^0_{\sigma;k}$ correspond to the elliptically polarized waves. For example, for the circularly polarized waves we have $A^0_{\sigma;k} \propto \left( I \pm i(I \times k)|k|^{-1} \right)/\sqrt{2}$, where $I$ is a real unit vector that is orthogonal to the wave vector $k$. Hence, obviously, we obtain $A^0_{\sigma;-k} = A^{0*}_{\sigma;k}$.

$$\hat{H} = \sum_{k,\sigma} \hbar c|k| \left( \frac{\hat{b}^{\dagger}_{\sigma;k}\hat{b}_{\sigma;k}}{K_k} + \frac{1}{2} \right) + V\sum_p W_p \hat{\rho}_{pp;0} + \left( 2\pi - \frac{\alpha}{2} \right) V \sum_k \hat{P}^{\dagger}_k \hat{P}_k + \hat{H}_{int}, \qquad (13a)$$

where

$$\hat{P} = \sum_k \hat{P}_k e^{ikr}, \quad \hat{P}_k = \frac{1}{V}\int_V \hat{P} e^{-ikr} d^3r, \quad \hat{P}_{-k} = \hat{P}^{\dagger}_k \qquad (13b)$$

are the spatial Fourier harmonics of the polarization operator and

$$\hat{H}_{int} = -V\sum_{k,\sigma}\left( \hat{b}_{\sigma;k} D^0_{\sigma;k} \hat{P}^{\dagger}_k + \hat{b}^{\dagger}_{\sigma;k} D^{0*}_{\sigma;k} \hat{P}_k \right) \qquad (13c)$$

is the operator of interaction between the "transverse" field and the medium.

The harmonics of the polarization operator are expressed through the harmonics of the density operator:

$$\hat{P}_k = \sum_{m,n} d_{nm}\hat{\rho}_{mn;k}, \quad \hat{\rho}_{mn;k} = \frac{1}{V}\int_V \hat{\rho}_{mn} e^{-ikr} d^3r,$$

$$\hat{\rho}_{mn} = \sum_k \hat{\rho}_{mn;k} e^{ikr}, \quad \hat{\rho}_{mn;-k} = \hat{\rho}^{\dagger}_{nm;k};$$

the operator $\hat{\rho}_{pp;0}$ in Eq. (13a) denotes $\hat{\rho}_{pp;k=0}$, i.e., $V\sum_p W_p \hat{\rho}_{pp;0} = \int_V \sum_p W_p \hat{\rho}_{pp}(r)d^3r$.

Making use of the Heisenberg equations for the field operators

$$\dot{\hat{b}}_{\sigma;k} = \frac{i}{\hbar}[\hat{H},\hat{b}_{\sigma;k}], \quad \dot{\hat{b}}^{\dagger}_{\sigma;k} = \frac{i}{\hbar}[\hat{H},\hat{b}^{\dagger}_{\sigma;k}],$$

we find

$$\dot{\hat{b}}_{\sigma;k} + ic|k|\hat{b}_{\sigma;k} = i\frac{VK_k}{\hbar}D^{0*}_{\sigma;k}\hat{P}_k, \quad \dot{\hat{b}}^{\dagger}_{\sigma;k} - ic|k|\hat{b}^{\dagger}_{\sigma;k} = -i\frac{VK_k}{\hbar}D^0_{\sigma;k}\hat{P}^{\dagger}_k. \qquad (14a)$$

Commutation relation for the harmonics of the density operator follows from Eq.(1):

$$[\hat{\rho}_{qp;k},\hat{\rho}_{mn;k'}] = \frac{1}{V}\left( \hat{\rho}_{mp;k+k'}\delta_{qn} - \delta_{mp}\hat{\rho}_{qn;k+k'} \right); \qquad (14c)$$

it can be verified that the Heisenberg equations

$$\dot{\hat{\rho}}_{mn;k} = \frac{i}{\hbar}[\hat{H},\hat{\rho}_{mn;k}]$$

lead to the Fourier images for relations (7a), (7b), and (7c).

# VI. CONNECTION BETWEEN A VACUUM PHOTON AND A PHOTON IN THE MEDIUM. BOGOLIUBOV TRANSFORMATION

The secondary quantization procedure applied to the phenomenological equations of the field in the medium [2,19] leads to the standard Heisenberg equations for the photon creation and annihilation operators [1,2,4,5,7,48], whose energy and momentum are related by a classical dispersion equation for the field in the medium. We will show that beyond the phenomenological theory the corresponding operators of creation and annihilation of "photons in the medium" result from the linear transformation (Bogoliubov transformation), which were applied to the "vacuum" photon creation and annihilation operators introduced in Section **V**. The parameter of this transformation is the refractive index of the wave considered, $n^2_{\sigma;k} = c^2 k^2 / \omega^2$.

Appropriate relations can be obtained most easily for the transversely polarized wave, whose description requires the use of the vector potential alone. The wave can be transverse (i.e., $\hat{\boldsymbol{E}}_\sigma \perp \boldsymbol{k}$) if its wave vector $\boldsymbol{k}$ is orthogonal to one of the eigenvectors $\boldsymbol{u}_\sigma$ of the tensor $\tilde{\tilde{\chi}}^a_\omega$ introduced above:

$$\tilde{\tilde{\chi}}^a_\omega \boldsymbol{u}_\sigma = \chi^a_{\sigma;\omega} \boldsymbol{u}_\sigma,$$

where $\chi^a_{\sigma;\omega}$ is the corresponding eigenvalue. Assuming[5] that $\boldsymbol{A}^0_{\sigma;k} \uparrow\uparrow \boldsymbol{u}_\sigma$, the tensors $\tilde{\tilde{\chi}}(r,\tau)$ and $\tilde{\tilde{\chi}}_\omega(r)$ used in Eqs. (10a) and (10b) can be replaced by the scalars $\chi_\sigma(\tau)$ and $\chi_{\sigma,\omega}$:

$$\hat{\boldsymbol{P}} = \hat{\chi}_\sigma \hat{\boldsymbol{E}} \equiv \int_0^\infty \chi_\sigma(\tau) \hat{\boldsymbol{E}}(r,t-\tau) d\tau, \tag{15a}$$

$$\int_0^\infty \chi_\sigma(\tau) e^{i\omega\tau} d\tau = \chi_{\sigma,\omega} = \frac{\chi^a_{\sigma;\omega}}{1 - \alpha \chi^a_{\sigma;\omega}}. \tag{15b}$$

We then take into account that the transverse electric field satisfies the relation $\hat{\boldsymbol{E}}_\sigma = -c^{-1} \dot{\hat{\boldsymbol{A}}}_\sigma$. Comparing the spectral expansion of the field operator $\hat{\boldsymbol{E}}_\sigma = \sum_{\sigma,k} \hat{\boldsymbol{E}}_{\sigma;k} \exp(i\boldsymbol{k}\boldsymbol{r})$ with expressions (11a) and (11c) for the vector potential, we arrive at the relation

$$\hat{\boldsymbol{E}}_{\sigma;k} = -\frac{\boldsymbol{A}^0_{\sigma;k}}{c} \left( \dot{\hat{b}}_{\sigma;k} + \dot{\hat{b}}^\dagger_{\sigma;-k} \right). \tag{16}$$

Using Eqs.(14a) for the creation and annihilation operators $\hat{b}_{\sigma;k}$ and $\hat{b}^\dagger_{\sigma;-k}$, and allowing for expressions (15a), (15b), (16), (12a), and (12c), we obtain the following system of equations:

$$\dot{\hat{b}}_{\sigma;k} + ic|\boldsymbol{k}|\hat{b}_{\sigma;k} = -2\pi \hat{\chi}_\sigma \left( \dot{\hat{b}}_{\sigma;k} + \dot{\hat{b}}^\dagger_{\sigma;-k} \right), \quad \dot{\hat{b}}^\dagger_{\sigma;-k} - ic|\boldsymbol{k}|\hat{b}^\dagger_{\sigma;-k} = -2\pi \hat{\chi}_\sigma \left( \dot{\hat{b}}_{\sigma;k} + \dot{\hat{b}}^\dagger_{\sigma;-k} \right). \tag{17a}$$

It is seen that while in vacuum the operators $\hat{b}_{\sigma;k}$ and $\hat{b}^\dagger_{\sigma;-k}$ are linearly independent solutions of the Heisenberg equations (i.e., $\hat{b}_{\sigma;k} \propto \exp(-ic|\boldsymbol{k}|t), \hat{b}^\dagger_{\sigma;-k} \propto \exp(+ic|\boldsymbol{k}|t)$), in the medium the pairs $\hat{b}_{\sigma;k}$ and $\hat{b}^\dagger_{\sigma;-k}$ (and, of course, $\hat{b}_{\sigma;-k}$ and $\hat{b}^\dagger_{\sigma;k}$) are coupled. To avoid a misunderstanding, we note that this relation does not indicate the scattering process, because without specifying the time depen-dence of the operators $\hat{b}_{\sigma;k}$, it is impossible to associate them with the waves propagating in definite directions!

---

[5] Rigorously speaking, the direction of the vector $\boldsymbol{u}_\sigma$ can depend on the frequency $\omega$. However, it does not matter since the final relations will be given for monochromatic fields.

For the harmonic processes ($\propto \exp(-i\omega t)$), Eqs.(17a) imply an algebraic system given by

$$\begin{pmatrix} 2\pi\omega\chi_{\sigma,\omega}+(\omega-c|k|) & 2\pi\omega\chi_{\sigma,\omega} \\ 2\pi\omega\chi_{\sigma,\omega} & 2\pi\omega\chi_{\sigma,\omega}+(\omega+c|k|) \end{pmatrix} \cdot \begin{pmatrix} \hat{b}_{\sigma;k} \\ \hat{b}^{\dagger}_{\sigma;-k} \end{pmatrix} = 0, \tag{17b}$$

which generates the dispersion equation

$$c^2 k^2 - \omega^2 n^2_{\sigma;k} = 0, \tag{17c}$$

where $n^2_{\sigma;k} = 1 + 4\pi\chi_{\sigma,\omega}$, such that we always have $\chi_{\sigma,\omega}(-\omega) = \chi_{\sigma,\omega}(\omega)$ for the transparent medium [23].

We emphasize that dispersion equation (17c) was obtained within the framework of a self-consistent quantum approach, i.e., without using the phenomenological equations of electrodynamics of a continuous medium.

Dispersion equation (17c) corresponds to two solutions, $\omega_{1,2} = \pm\omega_{\sigma;k}$, where $\omega_{\sigma;k} = \sqrt{c^2 k^2 / n^2_{\sigma;k}}$. The solution of Eq.(17a) can be represented in the following form:

$$\begin{pmatrix} \hat{b}_{\sigma;k} \\ \hat{b}^{\dagger}_{\sigma;-k} \end{pmatrix} = \frac{n_{\sigma;k}+1}{Q}\begin{pmatrix} \hat{c}_{\sigma;k}\,e^{-i\omega_{\sigma;k}t} \\ \hat{s}_{\sigma;k}\,e^{i\omega_{\sigma;k}t} \end{pmatrix} - \frac{n_{\sigma;k}-1}{Q}\begin{pmatrix} \hat{s}_{\sigma;k}\,e^{i\omega_{\sigma;k}t} \\ \hat{c}_{\sigma;k}\,e^{-i\omega_{\sigma;k}t} \end{pmatrix}, \tag{18a}$$

where $\hat{c}_{\sigma;k}$ and $\hat{s}_{\sigma;k}$ are time-independent (Schrödinger) operators, and $Q$ is the normalization coefficient. After the Hermitian conjugation in Eqs.(18a) and the replacement $k \to -k$, we obtain $\hat{s}_{\sigma;k} = \hat{c}^{\dagger}_{\sigma;-k}$. Allowing for the latter relation and choosing the normalization constant $Q = 2\sqrt{n_{\sigma;k}}$, one can represent solution (18a) in a more compact form,

$$\begin{pmatrix} \hat{b}_{\sigma;k} \\ \hat{b}^{\dagger}_{\sigma;-k} \end{pmatrix} = \mathrm{ch}\,\zeta \begin{pmatrix} \hat{q}_{\sigma;k} \\ \hat{q}^{\dagger}_{\sigma;-k} \end{pmatrix} - \mathrm{sh}\,\zeta \begin{pmatrix} \hat{q}^{\dagger}_{\sigma;-k} \\ \hat{q}_{\sigma;k} \end{pmatrix}, \tag{18b}$$

where $\zeta = \ln n^2_{\sigma;k}/4$, and the operators $\hat{q}_{\sigma;k} = \hat{c}_{\sigma;k}\,e^{-i\omega_k t}$ and $\hat{q}^{\dagger}_{\sigma;k} = \hat{c}^{\dagger}_{\sigma;k}\,e^{i\omega_k t}$ are described by standard equations for the Heisenberg creation and annihilation operators:

$$\left.\begin{array}{ll} \dot{\hat{q}}_{\sigma;k} = -i\omega_{\sigma;k}\hat{q}_{\sigma;k}, & \dot{\hat{q}}^{\dagger}_{\sigma;k} = i\omega_{\sigma;k}\hat{q}^{\dagger}_{\sigma;k} \\ \dot{\hat{q}}_{\sigma;-k} = -i\omega_{\sigma;k}\hat{q}_{\sigma;-k}, & \dot{\hat{q}}^{\dagger}_{\sigma;-k} = i\omega_{\sigma;k}\hat{q}^{\dagger}_{\sigma;-k} \end{array}\right\}. \tag{18c}$$

Thus, dispersion equation (17c) corresponds to the operators related to the "vacuum" operators $\hat{b}_{\sigma;k}$ and $\hat{b}^{\dagger}_{\sigma;-k}$ by the Bogoliubov transformation [31, 32], which is inverse of (18b):

$$\begin{pmatrix} \hat{q}_{\sigma;k} \\ \hat{q}^\dagger_{\sigma;-k} \end{pmatrix} = \mathrm{ch}\,\zeta \begin{pmatrix} \hat{b}_{\sigma;k} \\ \hat{b}^\dagger_{\sigma;-k} \end{pmatrix} + \mathrm{sh}\,\zeta \begin{pmatrix} \hat{b}^\dagger_{\sigma;-k} \\ \hat{b}_{\sigma;k} \end{pmatrix}. \tag{19a}$$

Operators $\hat{q}_{\sigma;k}$ and $\hat{q}^\dagger_{\sigma;k}$ satisfy the same commutation relations as the operators $\hat{b}_{\sigma;k}$ and $\hat{b}^\dagger_{\sigma;k}$ of the "vacuum" field. In the limiting case of vacuum ($n^2_{\sigma;k}=1$) we obtain $\hat{q}_{\sigma;k} \equiv \hat{b}_{\sigma;k}$. It can easily be verified that relations (19a) are equivalent to the relation between complex amplitudes of the classical transverse electromagnetic waves at the medium/vacuum interface at normal incidence.

Neglecting the relation between the vacuum operators $\hat{b}_{\sigma;k}$ and $\hat{b}^\dagger_{\sigma;-k}$, which appears in the case of the field in the medium, i.e., assuming in Eq.(14a) that $\hat{P}_{\sigma;k} = -\dfrac{A^0_{\sigma;k}}{c}\hat{\chi}_\sigma \dot{\hat{b}}_{\sigma;k}$, $\hat{P}^\dagger_{\sigma;k} = -\dfrac{A^{0*}_{\sigma;k}}{c}\hat{\chi}_\sigma \dot{\hat{b}}^\dagger_{\sigma;k}$, instead of (17a) we would obtained the equations

$$(1+2\pi\hat{\chi}_\sigma)\dot{\hat{b}}_{\sigma;k} + ic|k|\hat{b}_{\sigma;k} = 0, \quad (1+2\pi\hat{\chi}_\sigma)\dot{\hat{b}}^\dagger_{\sigma;k} - ic|k|\hat{b}^\dagger_{\sigma;k} = 0,$$

from which the approximate dispersion relation $\omega(1+2\pi\chi_{\sigma,\omega}) = \pm c|k|$ follows. This is exactly the approximation of small optical density to which the widespread approach discussed in the Introduction (see, e.g., [24, 25]) corresponds.

Using relations (16) and (18b), we arrive at the following representation of the operator of the transverse field in the medium:

$$\hat{E}_\sigma = \sum_k \left( E^0_{\sigma;k}\hat{q}_{\sigma;k}\,\mathrm{e}^{ikr} + E^{0*}_{\sigma;k}\hat{q}^\dagger_{\sigma;k}\,\mathrm{e}^{-ikr} \right) = \sum_k \left( E^0_{\sigma;k}\hat{c}_{\sigma;k}\,\mathrm{e}^{-i\omega_{\sigma;k}t+ikr} + E^{0*}_{\sigma;k}\hat{c}^\dagger_{\sigma;k}\,\mathrm{e}^{i\omega_{\sigma;k}t-ikr} \right), \tag{19b}$$

where $E^0_{\sigma;k}$ is the normalized vector, $E^0_{\sigma;k} \uparrow\uparrow u_\sigma \perp k$.

In the case of arbitrary orientation of the wave vector $k$, normal waves in the anisotropic medium are not transverse in general [23,49]. Electromagnetic field is specified by the relation which follows from Eqs. (6b), (11a), and (11c):

$$\hat{E} = \sum_{\sigma,k} \hat{E}_{\sigma;k}\,\mathrm{e}^{ikr}, \quad \hat{E}_{\sigma;k} = -\dfrac{A^0_{\sigma;k}}{c}\left(\dot{\hat{b}}_{\sigma;k} + \dot{\hat{b}}^\dagger_{\sigma;-k}\right) - 4\pi\hat{P}_{\|\sigma;k},$$

where $\hat{P}_{\|\sigma;k} \uparrow\uparrow k$. The longitudinal component of the field should be proportional to the transverse one since the condition $\ddot{\varepsilon}\hat{E}_{\sigma;k} \perp k$ follows from Eqs. (11b), (12c), and (10c). Finally, we arrive at the following expression:

$$\hat{E}_{\sigma;k} = -iE^0_{\sigma;k}\left(\dot{\hat{b}}_{\sigma;k} + \dot{\hat{b}}^\dagger_{\sigma;-k}\right), \tag{20a}$$

where the normalization vector $E^0_{\sigma;k}$ (which we have to determine) is, in general, not orthogonal to the wave vector $k$. The orientation of the vector $E^0_{\sigma;k}$ can most easily be found using Eqs. (10c) and (10d). We will write these vector operator equations in the coordinate system whose axis Z is directed along the wave vector $\boldsymbol{k}$. Assuming the harmonic temporal dependence of the operators $\hat{b}_{\sigma;\boldsymbol{k}}, \hat{b}^{\dagger}_{\sigma;-\boldsymbol{k}} \propto \exp(-i\omega t)$, we obtain

$$\left(\vec{\varepsilon}_{\omega} - \vec{1}_{\perp} n^2_{\sigma;\boldsymbol{k}}\right)\boldsymbol{E}^0_{\sigma;\boldsymbol{k}} = 0, \tag{20b}$$

where $\vec{1}_{\perp}$ is the diagonal tensor, in which the elements "$xx$" and "$yy$" are equal to 1, and the element "$zz$" is equal to zero. Relation (20b) determines both the orientation of the vectors $\boldsymbol{E}^0_{\sigma;\boldsymbol{k}}$ and the refractive indices $n^2_{\sigma;\boldsymbol{k}}$ of the normal waves. The latter are the roots of the biquadratic equation

$$\det\left(\vec{\varepsilon}_{\omega} - \vec{1}_{\perp} n^2_{\sigma;\boldsymbol{k}}\right) = 0. \tag{20c}$$

Using Eqs. (14a) and (20b), after simple (but rather cumbersome) transformations we obtain an algebraic system of equations for the operators $\hat{b}_{\sigma;\boldsymbol{k}}$ and $\hat{b}^{\dagger}_{\sigma;-\boldsymbol{k}}$. The corresponding system exactly coincides with Eqs.(17b) after the replacement

$$2\pi\omega\chi_{\sigma,\omega} \Rightarrow \frac{n^2_{\sigma;\boldsymbol{k}} - 1}{2},$$

where as the squared quadratic indices $n^2_{\sigma;\boldsymbol{k}}$ of the normal waves, the solutions of Eq.(20c) should be used. It can easily be verified that relations (18b), (18c), (19a), and (19b) remain valid in this case, too, if one has in mind the refractive indices $n^2_{\sigma;\boldsymbol{k}}$ and the vectors $\boldsymbol{E}^0_{\sigma;\boldsymbol{k}}$ determined in the way mentioned above. As a result, the field operator is given by an expression that generalizes Eq. (19b):

$$\hat{\boldsymbol{E}} = \sum_{\sigma,\boldsymbol{k}}\left(\boldsymbol{E}^0_{\sigma;\boldsymbol{k}}\hat{q}_{\sigma;\boldsymbol{k}}\, e^{i\boldsymbol{k}\boldsymbol{r}} + \boldsymbol{E}^{0*}_{\sigma;\boldsymbol{k}}\hat{q}^{\dagger}_{\sigma;\boldsymbol{k}}\, e^{-i\boldsymbol{k}\boldsymbol{r}}\right) = \sum_{\sigma,\boldsymbol{k}}\left(\boldsymbol{E}^0_{\sigma;\boldsymbol{k}}\hat{c}_{\sigma;\boldsymbol{k}}\, e^{-i\omega_{\sigma;\boldsymbol{k}}t + i\boldsymbol{k}\boldsymbol{r}} + \boldsymbol{E}^{0*}_{\sigma;\boldsymbol{k}}\hat{c}^{\dagger}_{\sigma;\boldsymbol{k}}\, e^{i\omega_{\sigma;\boldsymbol{k}}t - i\boldsymbol{k}\boldsymbol{r}}\right). \tag{21}$$

If the tensor components $\varepsilon^{(xz)}_{\omega} = \varepsilon^{(zx)*}_{\omega}$ and $\varepsilon^{(yz)}_{\omega} = \varepsilon^{(zy)*}_{\omega}$ are equal to zero in the considered coordinate system (i.e., where the Z axis is directed along the vector vector $\boldsymbol{k}$), then we pass to the transverse waves considered above. However, in the indicated "degenerate" case, system (20b) splits into independent equations for the transverse and longitudinal (with respect to the wave vector $\boldsymbol{k}$) waves, which correspond to the equation $\varepsilon^{(zz)}_{\omega}\hat{\boldsymbol{E}}_{\|\boldsymbol{k}} = 0$. The longitudinal waves are purely potential: $\hat{\boldsymbol{E}}_{\|\boldsymbol{k}} = -4\pi\hat{\boldsymbol{P}}_{\|\boldsymbol{k}} = i\boldsymbol{k}\hat{\varphi}_{\boldsymbol{k}}$. Potential waves satisfy the dispersion equation $\varepsilon^{(zz)}_{\omega}(\omega_{\boldsymbol{k}}) = 0$, which is similar to that for the Langmuir wave in a plasma. The secondary quantization procedure applied to the potential oscillations leads to a relation similar to (21). Thus, expression (21) can always be used by associating different values of the index $\sigma$ to all independent solutions of Eq.(20b). There are three such solutions in the case $\varepsilon^{(xy)}_{\omega} = \varepsilon^{(xz)}_{\omega} = 0$. We emphasize that while for the waves with a nonzero vortex field component the quantum creation and annihilation operators of the field in the medium are obtained by applying the Bogoliubov transformation to the "vacuum" operators, the operators of purely potential waves and the vacuum operators in a homogeneous medium are not connected.

It is interesting to follow the results of applying the approach used above to the fields expanded over real orthogonal modes $A_\alpha(r)$ (in particular, for the fields in the resonant cavities). In this case, instead of Eq.(11a) we have $\hat{A} = \sum_\alpha (\hat{b}_\alpha + \hat{b}_\alpha^\dagger) A_\alpha(r)$. Let the mode with the spatial structure $A_\alpha(r)$ in vacuum correspond to the eigenfrequency $\omega_\alpha^0$. Then, using the same algebra as in the case of expansion of the field over complex harmonics, we obtain

$$\left. \begin{array}{l} \hat{A} = \sum_\alpha (\hat{q}_\alpha + \hat{q}_\alpha^\dagger) \tilde{A}_\alpha(r) \\ \dot{\hat{q}}_\alpha = -i\omega_\alpha \hat{q}_\alpha, \quad \dot{\hat{q}}_\alpha^\dagger = i\omega_\alpha \hat{q}_\alpha^\dagger \end{array} \right\},$$

where $\omega_\alpha = \omega_\alpha^0/n_\alpha$, $n_\alpha^2 = 1 + 4\pi\chi_\alpha(\omega_\alpha)$, and $\chi_\alpha(\omega_\alpha)$ is the polarizability of the medium; $\tilde{A}_\alpha(r)$ are the renormalized modes amplitudes; the operators $\hat{q}_\alpha$ and $\hat{q}_\alpha^\dagger$ are connected with the "vacuum" operators $\hat{b}_\alpha$ and $\hat{b}_\alpha^\dagger$ by the same Bogoliubov transform [31, 32] with the parameter $\zeta = \ln n_\alpha^2 / 4$:

$$\left. \begin{array}{l} \hat{q}_\alpha = \hat{b}_\alpha \operatorname{ch} \zeta + \hat{b}_\alpha^\dagger \operatorname{sh} \zeta \\ \hat{q}_\alpha^\dagger = \hat{b}_\alpha^\dagger \operatorname{ch} \zeta + \hat{b}_\alpha \operatorname{sh} \zeta \end{array} \right\}.$$

The latter relation was also obtained in [15] for particular case of a resonator with metal walls filled by homogeneous dielectric without frequency dispersion. It can easily be verified that in the case of expansion over harmonic functions $A_{\alpha,\alpha'}(r) \propto \cos kr, \sin kr$, from the latter relation one can obtain expressions (19a) and (19b).

VII. PHOTON ENERGY IN THE MEDIUM

The choice of the magnitudes of the normalization vectors $E_{\sigma;k}^0$ in Eq. (21) is not important when the linear problems with initial or boundary conditions are addressed. However, having in mind the problems of generation, detection, and nonlinear interaction of waves, it is convenient to determine the vectors $E_{\sigma;k}^0$ in such a way that energy variation of the "medium+field" system with the field "switching on" is described by the operator $\hat{H}$ of standard form:

$$\hat{H} = \sum_{k,\sigma} \hbar\omega_{\sigma;k} \frac{1}{2} (\hat{q}_{\sigma;k}^\dagger \hat{q}_{\sigma;k} + \hat{q}_{\sigma;k} \hat{q}_{\sigma;k}^\dagger) = \sum_{k,\sigma} \hbar\omega_{\sigma;k} \left( \hat{q}_{\sigma;k}^\dagger \hat{q}_{\sigma;k} + \frac{1}{2} \right). \tag{22a}$$

If so, the Heisenberg equation

$$\dot{\hat{q}}_{\sigma;k} = \frac{i}{\hbar}[\hat{H}, \hat{q}_{\sigma;k}], \quad \dot{\hat{q}}_{\sigma;k}^\dagger = \frac{i}{\hbar}[\hat{H}, \hat{q}_{\sigma;k}^\dagger] \tag{22b}$$

will correspond to Eqs.(18c).

Within the framework of phenomenological theory [2,19], appropriate normalization is chosen by using an expression determining the energy of the classical wave field of the form $E = \sum_{\sigma,k} E_{\sigma;k} \exp(i k r - i\omega_{\sigma;k} t) + c.c.$ in the medium [2,19,23,49]. The transition to the quantum-mechanical Hamiltonian is accomplished by replacing the classical quantity $2|E_{\sigma;k}|^2$ with the operator $2|E_{\sigma;k}|^2 \Rightarrow |E^0_{\sigma;k}|^2 (\hat{q}^\dagger_{\sigma;k}\hat{q}_{\sigma;k} + \hat{q}_{\sigma;k}\hat{q}^\dagger_{\sigma;k})$ (see [2,19,20]). We obtain the corresponding normalization condition of the field without using a phenomenological approach. For this purpose, we use the expression for quantum energy operator (2a), from which we subtract the energy of the undisturbed medium:

$$\hat{H} = \int_V \left( \frac{\hat{E}^2 + \hat{B}^2}{8\pi} - \frac{\alpha}{2}\hat{P}^2 + \hat{w} \right) d^3 r - \int_V \sum_m N_m(r) W_m d^3 r, \qquad (22c)$$

where the spatial distributions $N_m(r)$ correspond to the undisturbed medium. For the energy density operator $\hat{w}$ of the internal degrees of freedom we use a formula that follows from the equations of motion (7a), (7b), and (7c) and definitions (2b), (2c):

$$\hat{w} = \sum_m N_m(r) W_m + \frac{\alpha}{2}\hat{P}^2 + \frac{1}{4\pi} \int_{-\infty}^{t} (\dot{\hat{D}}\hat{E}) d\tau - \frac{\hat{E}^2}{8\pi}. \qquad (23a)$$

Then we allow for linear connection (10c, d) between the operators $\hat{D}$ and $\hat{E}$. Given that relations (10c, d) are identical to the corresponding classical expressions, one can make use of the technique described in [2,19,23]. As a result, we obtain

$$\int_V \int_{-\infty}^{t} (\dot{\hat{D}}\hat{E}) d\tau d^3 r = \frac{V}{2} \sum_{k,\sigma} \left[ E^{0*}_{\sigma;k} \cdot \left.\frac{\partial(\omega\vec{\varepsilon}_\omega)}{\partial \omega}\right|_{\omega=\omega_{\sigma;k}} \cdot E^0_{\sigma;k} \right] (\hat{q}^\dagger_{\sigma;k}\hat{q}_{\sigma;k} + \hat{q}_{\sigma;k}\hat{q}^\dagger_{\sigma;k}). \qquad (23b)$$

The magnetic field operator is given by

$$\hat{B} = \sum_{k,\sigma} \frac{n_{\sigma;k}}{|k|} (k \times E^0_{\sigma;k}) \hat{q}_{\sigma;k} e^{i k r} + H.c.;$$

Using Eq.(20b), the expression for the operator $\int_V \hat{B}^2 d^3 r$ can be transformed to the following form:

$$\int_V \hat{B}^2 d^3 r = V \sum_{k,\sigma} [E^{0*}_{\sigma;k} \cdot \vec{\varepsilon}_\omega|_{\omega=\omega_{\sigma;k}} \cdot E^0_{\sigma;k}](\hat{q}^\dagger_{\sigma;k}\hat{q}_{\sigma;k} + \hat{q}_{\sigma;k}\hat{q}^\dagger_{\sigma;k}). \qquad (23c)$$

In view of expressions (22c), (23a), (23b), and (23c), we obtain

$$\hat{H} = \frac{V}{4\pi} \sum_{k,\sigma} \left[ E^{0*}_{\sigma;k} \cdot \left.\frac{\partial(\omega^2 \vec{\varepsilon}_\omega)}{\partial \omega^2}\right|_{\omega=\omega_{\sigma;k}} \cdot E^0_{\sigma;k} \right] (\hat{q}^\dagger_{\sigma;k}\hat{q}_{\sigma;k} + \hat{q}_{\sigma;k}\hat{q}^\dagger_{\sigma;k}). \qquad (24a)$$

If the condition $E^{0*}_{\sigma;k}(\partial(\omega^2\vec{\varepsilon}_\omega)/\partial\omega^2)E^0_{\sigma;k} > 0$ is fulfilled, then the normalization condition leading to the standard form (22a) of the Hamiltonian follows from Eq.(24a):

$$\left[\boldsymbol{E}_{\sigma;k}^{0*} \cdot \frac{\partial(\omega^2 \vec{\vec{\varepsilon}}_{\omega})}{\partial \omega^2}\bigg|_{\omega=\omega_{\sigma;k}} \cdot \boldsymbol{E}_{\sigma;k}^{0}\right] = \frac{2\pi\hbar\omega_{\sigma;k}^2}{V} \qquad (24b)$$

(the case $\boldsymbol{E}_{\sigma;k}^{0*}\left(\partial(\omega^2\vec{\vec{\varepsilon}}_{\omega})/\partial\omega^2\right)\boldsymbol{E}_{\sigma;k}^{0} < 0$ is considered below). Thus, Hamiltonian (22a) corresponds to the energy of collective excitation of the field and the medium, which includes the energy of the macroscopic electromagnetic field, the energy of the internal degrees of freedom, and the energy of the near-field dipole interaction in the polarized medium. In the limiting case of vacuum ($\vec{\vec{\varepsilon}} = \vec{\vec{1}}$) normalization (24b) has the standard form [1,4,48]

$$\left|\boldsymbol{E}_{\sigma;k}^{0}\right|^2 = \frac{2\pi\hbar\omega_{\sigma;k}}{V} = \frac{2\pi\hbar c|\boldsymbol{k}|}{V}.$$

## VIII. THE NEGATIVE ENERGY WAVES

The expression $\boldsymbol{E}_{\sigma;k}^{0*}\left(\partial(\omega^2\vec{\vec{\varepsilon}}_{\omega})/\partial\omega^2\right)\boldsymbol{E}_{\sigma;k}^{0}$ may be negative in some frequency range. It follows from expression (24a) that under the condition $\boldsymbol{E}_{\sigma;k}^{0*}\left(\partial(\omega^2\vec{\vec{\varepsilon}}_{\omega})/\partial\omega^2\right)\boldsymbol{E}_{\sigma;k}^{0} < 0$ the wave mode excitation is accompanied by a decrease in the energy of the "wave + medium" system, and such waves are therefore potentially unstable during interaction with another subsystem.[6] Dynamics of such systems with "excess" free energy can conveniently be described using the notion of negative energy. Within the framework of phenomenological theory under the condition $\boldsymbol{E}_{\sigma;k}^{0*}\left(\partial(\omega^2\vec{\vec{\varepsilon}}_{\omega})/\partial\omega^2\right)\boldsymbol{E}_{\sigma;k}^{0} < 0$ a model "negative" Hamiltonian is introduced for the wave mode:

$$\hat{H} = -\hbar\omega_{\sigma;k}\hat{q}_{\sigma;k}^{\dagger}\hat{q}_{\sigma;k} \qquad (25a)$$

(to avoid cumbersome expressions, we do not sum up over different spatial harmonics and polarizations of the field). The best known example of such a system are waves in the ensemble of inverted two-level atoms with a sufficiently high density (in this regard, see, e.g., [20] and references therein).

Although the sign on the right-hand side of Eq. (25a) is absolutely not critical in the description of stationary oscillations, the use of "negative" Hamiltonian (25a) in Eqs.(22b) with the standard commutation relation $[\hat{q}_{\sigma;k}, \hat{q}_{\sigma';k'}^{\dagger}] = \delta_{\sigma\sigma'}\delta_{kk'}$ must be accompanied by a cyclic permutation of the creation and annihilation operators in the expression for the field operator (21):

$$\hat{\boldsymbol{E}} = \boldsymbol{E}_{\sigma;k}^{0*}\hat{q}_{\sigma;k}^{\dagger} e^{i\boldsymbol{k}\boldsymbol{r}} + \boldsymbol{E}_{\sigma;k}^{0}\hat{q}_{\sigma;k} e^{-i\boldsymbol{k}\boldsymbol{r}} = \boldsymbol{E}_{\sigma;k}^{0*}\hat{c}_{\sigma;k}^{\dagger} e^{-i\omega_{\sigma;k}t+i\boldsymbol{k}\boldsymbol{r}} + \boldsymbol{E}_{\sigma;k}^{0}\hat{c}_{\sigma;k} e^{i\omega_{\sigma;k}t-i\boldsymbol{k}\boldsymbol{r}} \qquad (25b)$$

(exactly such a description was used in [20] within the framework of the phenomenological approach). We obtain an expression for the Hamiltonian of the wave mode in the case $\boldsymbol{E}_{\sigma;k}^{0*}\left(\partial\left(\omega^2\vec{\vec{\varepsilon}}_\omega\right)/\partial\omega^2\right)\boldsymbol{E}_{\sigma;k}^{0} < 0$ using the above-formulated rigorous equations of dynamics of the quantum field in the medium.

---

[6] For example, with waves of another type or with a dissipative basin [20].

Represent the polarization vector $\hat{\boldsymbol{P}}$ of the medium in the following form:

$$\hat{\boldsymbol{P}} = \vec{\vec{\chi}}\hat{\boldsymbol{E}} + \delta\hat{\boldsymbol{P}}, \tag{26a}$$

where $\vec{\vec{\chi}}$ is the linear operator of polarizability of the medium, which was introduced in Section **IV**, $\delta\hat{\boldsymbol{P}}$ is the "additional" polarization which allows for, e.g., the nonlinear interaction or the presence of external sources (e.g., classical pump) [2,37,42]. In this case, operator equation (10c) can be modified as follows:

$$\frac{\partial^2}{\partial t^2}\left(\vec{\vec{\varepsilon}}\cdot\hat{\boldsymbol{E}}\right) + c^2\nabla\times\nabla\times\hat{\boldsymbol{E}} = -4\pi\delta\ddot{\hat{\boldsymbol{P}}}, \tag{26b}$$

where $\vec{\vec{\varepsilon}} = 1 + 4\pi\vec{\vec{\chi}}$. Confining oneself to the spatially homogeneous problem, it suffices to consider one spatial harmonic of the operator $\delta\hat{\boldsymbol{P}}$:

$$\delta\hat{\boldsymbol{P}} = \delta\hat{\boldsymbol{P}}_k\,\mathrm{e}^{ikr} + \delta\hat{\boldsymbol{P}}_k^\dagger\,\mathrm{e}^{-ikr}. \tag{26c}$$

Furthermore, assuming that the disturbance of the polarization of $\delta\hat{\boldsymbol{P}}$ is, in a sense, small, we consider the resonant approximation:[7]

$$\delta\hat{\boldsymbol{P}}_k = \delta\hat{\boldsymbol{P}}_{0k}(t)\mathrm{e}^{-i\omega_{\sigma;k}t}, \tag{26d}$$

where the frequency $\omega_{\sigma;k}$ corresponds to the discussed wave mode, $\delta\dot{\hat{\boldsymbol{P}}}_{0k} \ll \omega_{\sigma;k}\hat{\boldsymbol{P}}_{0k}$.

We will seek the electromagnetic field in the form of the wave disturbance of type (21), but with a slowly time-dependent envelope [2,24,25,42,37]:

$$\hat{\boldsymbol{E}} = \boldsymbol{E}_{\sigma;k}^{0}\hat{q}_{\sigma;k}\,\mathrm{e}^{ikr} + H.C. = \boldsymbol{E}_{\sigma;k}^{0}\hat{c}_{\sigma;k}(t)\mathrm{e}^{-i\omega_{\sigma;k}t+ikr} + H.C. \tag{27a}$$

Then we use an auxiliary relation that is valid for the quasi-monochromatic field with a "slow" variable:[8] $\boldsymbol{E} = \boldsymbol{E}c(t)\exp(-i\omega t)$, where $\boldsymbol{E}$ is the time-independent vector. The corresponding expression in the form convenient for us is given in, e.g., [2,52]:

$$\mathrm{e}^{i\omega t}\cdot\boldsymbol{E}^*\frac{\partial^2\vec{\vec{\varepsilon}}\left(\boldsymbol{E}c(t)\mathrm{e}^{-i\omega t}\right)}{\partial t^2} \approx -\omega^2\left(\boldsymbol{E}^*\vec{\vec{\varepsilon}}_\omega\boldsymbol{E}\right)c(t) - 2i\omega\left[\boldsymbol{E}^*\frac{\partial\left(\omega^2\vec{\vec{\varepsilon}}_\omega\right)}{\partial\omega^2}\boldsymbol{E}\right]\frac{\partial c}{\partial t}. \tag{27b}$$

Substitute expressions (26c), (26d), and (27a) into Eq.(26b). We then multiply the resulting expression on the left by the factor $\boldsymbol{E}_{\sigma;k}^{0*}\exp(i\omega_{\sigma;k}t - ikr)$. Next, we allow for the formal inequality $\partial/\partial t \ll \omega_{\sigma;k}$ in all the relations, use expressions (27b) and (20b), and integrate over the volume $V$. Finally, we arrive at the following equations for the operator $\hat{c}_{\sigma;k}(t)$ of the envelope:

$$\Lambda_{\omega_{\sigma;k}}\dot{\hat{c}}_{\sigma;k} = 4\pi i\omega_{\sigma;k}^2\delta\hat{\boldsymbol{P}}_{0k}\boldsymbol{E}_{\sigma;k}^{0*}, \tag{28a}$$

---

[7] This case is typical, for example, of the polarization disturbance due to the parametric interaction of waves [2,,37,42,50,51].
[8] In this case, it does not matter whether the quantity "$c$" is an ordinary function or an operator.

where

$$\Lambda_{\omega_{\sigma;k}} = 2\omega_{\sigma;k}\left[\boldsymbol{E}^{0*}_{\sigma;k}\cdot\frac{\partial(\omega^2\vec{\varepsilon}_\omega)}{\partial\omega^2}\bigg|_{\omega=\omega_{\sigma;k}}\cdot\boldsymbol{E}^0_{\sigma;k}\right] = 2\omega_{\sigma;k}\eta_{\omega_{\sigma;k}}\left|\left[\boldsymbol{E}^{0*}_{\sigma;k}\cdot\frac{\partial(\omega^2\vec{\varepsilon}_\omega)}{\partial\omega^2}\bigg|_{\omega=\omega_{\sigma;k}}\cdot\boldsymbol{E}^0_{\sigma;k}\right]\right|,$$

$$\eta_{\omega_{\sigma;k}} = \pm 1 = \text{sign}\left[\boldsymbol{E}^{0*}_{\sigma;k}\cdot\frac{\partial(\omega^2\vec{\varepsilon}_\omega)}{\partial\omega^2}\bigg|_{\omega=\omega_{\sigma;k}}\cdot\boldsymbol{E}^0_{\sigma;k}\right].$$

We choose the field normalization which is similar to Eq.(24a), but allows for the possibility that the $\boldsymbol{E}^{0*}_{\sigma;k}(\partial(\omega^2\vec{\varepsilon}_\omega)/\partial\omega^2)\boldsymbol{E}^0_{\sigma;k}$ can be negative:

$$\left|\left[\boldsymbol{E}^{0*}_{\sigma;k}\cdot\frac{\partial(\omega^2\vec{\varepsilon}_\omega)}{\partial\omega^2}\bigg|_{\omega=\omega_{\sigma;k}}\cdot\boldsymbol{E}^0_{\sigma;k}\right]\right| = \frac{2\pi\hbar\omega^2_{\sigma;k}}{V}\,; \tag{28b}$$

as a result, we transform Eq.(28a) to the following form:

$$\eta_{\omega_{\sigma;k}}\dot{\hat{c}}_{\sigma;k} = i\frac{V}{\hbar}\left(\delta\hat{\boldsymbol{P}}_{0k}\cdot\boldsymbol{E}^{0*}_{\sigma;k}\right), \tag{28c}$$

and then obtain equations for the operators $\hat{q}_{\sigma;k}$ and $\hat{q}^\dagger_{\sigma;k}$:

$$\eta_{\omega_{\sigma;k}}\left(\dot{\hat{q}}_{\sigma;k} + i\omega_{\sigma;k}\hat{q}_{\sigma;k}\right) = i\frac{V}{\hbar}\left(\delta\hat{\boldsymbol{P}}_k\cdot\boldsymbol{E}^{0*}_{\sigma;k}\right), \quad \eta_{\omega_{\sigma;k}}\left(\dot{\hat{q}}^\dagger_{\sigma;k} - i\omega_{\sigma;k}\hat{q}^\dagger_{\sigma;k}\right) = -i\frac{V}{\hbar}\left(\delta\hat{\boldsymbol{P}}^\dagger_k\cdot\boldsymbol{E}^0_{\sigma;k}\right) \tag{28d}$$

(for the specific case $\eta_{\omega_{\sigma;k}} = +1$ the equations of type (28d) were obtained in [2] within the framework of phenomenological theory). Representing Eqs.(28d) as the Heisenberg equations of form (22b), we obtain an expression for the corresponding Hamiltonian:

$$\hat{H} = \hbar\omega_{\sigma;k}\hat{q}^\dagger_{\sigma;k}\hat{q}_{\sigma;k} + \eta_{\omega_{\sigma;k}}\delta\hat{H}, \tag{29a}$$

where

$$\delta\hat{H} = -V\left(\delta\hat{\boldsymbol{P}}_k\boldsymbol{E}^{0*}_{\sigma;k}\hat{q}^\dagger_{\sigma;k} + \delta\hat{\boldsymbol{P}}^\dagger_k\boldsymbol{E}^0_{\sigma;k}\hat{q}_{\sigma;k}\right) \tag{29b}$$

is the operator of interaction of the wave mode with another subsystem (subsystems) written in the electric-dipole approximation. Strictly speaking, for a self-contained description, Hamiltonian (29a) should be supplemented with an "intrinsic" energy operator of the mentioned subsystem (subsystems). However, for the reasoning suggested here it suffices to discuss a simple model of independent external pump that specifies the "external" polarization $\delta\hat{\boldsymbol{P}}$.

It is evident that the interaction operator enters the Hamiltonian of the system with different signs, depending on the value of the parameter $\eta_{\omega_{\sigma;k}} = \pm 1$. However, the sign of the interaction operator can be retained by changing the sign of the wave mode energy. To this end, as was already mentioned, the wave field in the case $\eta_{\omega_{\sigma;k}} = -1$ should be represented in the form (25b). Thus, we obtain the relations

$$\hat{H} = -\hbar\omega_{\sigma;k}\hat{q}^\dagger_{\sigma;k}\hat{q}_{\sigma;k} + \delta\hat{H}, \tag{30a}$$

$$\delta\hat{H} = -V\left(\delta\hat{\boldsymbol{P}}_k \boldsymbol{E}^0_{\sigma;k} \hat{q}_{\sigma;k} + \delta\hat{\boldsymbol{P}}^\dagger_k \boldsymbol{E}^{0*}_{\sigma;k} \hat{q}^\dagger_{\sigma;k}\right) \tag{30b}$$

(changes in the form of the field presentation resulted in a corresponding cyclic permutation of the creation and annihilation operators in the expression for the operator $\delta\hat{H}$ !). Obviously, the use of relations (27a), (29a), and (29b) for any value of the parameter $\eta_{\omega_{\sigma;k}} = \pm 1$ or the transition to expressions (25b), (30a), and (30b) for $\eta_{\omega_{\sigma;k}} = -1$ is only is a matter of convenience since the equations of dynamics for "slow" amplitudes of the wave mode remain the same!

We can offer a more economical way to represent the Hamiltonian. Namely, we change the sign of the energy field depending on the sign of the quantity $\boldsymbol{E}^{0*}_{\sigma;k}\left(\partial(\omega^2 \ddot{\varepsilon}_\omega)/\partial\omega^2\right)\boldsymbol{E}^0_{\sigma;k}$, retaining both the form of representation of the field (27a) and the form of interaction operator (29b). It suffices to set the normalization of the commutation relation for the operators $\hat{q}_{\sigma;k}, \hat{q}^\dagger_{\sigma';k'}$ in accordance with the sign of the expression $\eta_{\omega_{\sigma;k}} = \pm 1$:

$$[\hat{q}_{\sigma;k}, \hat{q}^\dagger_{\sigma';k'}] = \delta_{\sigma\sigma'}\delta_{kk'}\eta_{\omega_{\sigma;k}}. \tag{31a}$$

In this case, Eqs. (28d) in the form (22b) correspond to the Hamiltonian

$$\hat{H} = \eta_{\omega_{\sigma;k}}\hbar\omega_{\sigma;k}\hat{q}^\dagger_{\sigma;k}\hat{q}_{\sigma;k} + \delta\hat{H}, \tag{31b}$$

where the operator $\delta\hat{H}$ is determined by Eq. (29b).

It is useful to give an equation describing the evolution of the photon number operator $\hat{N}_{\sigma;k} = \hat{q}^\dagger_{\sigma;k}\hat{q}_{\sigma;k}$. This expression has a particularly clear physical meaning if one introduces the operators of the negative frequency and positive frequency components of the "external" resonant current:

$$\delta\hat{\boldsymbol{J}} = \delta\dot{\hat{\boldsymbol{P}}} = \delta\hat{\boldsymbol{J}}_k e^{ikr} + \delta\hat{\boldsymbol{J}}^\dagger_k e^{-ikr},$$

where $\delta\hat{\boldsymbol{J}}_k = -i\omega_{\sigma;k}\delta\hat{\boldsymbol{P}}_k$, $\delta\hat{\boldsymbol{J}}^\dagger_k = i\omega_{\sigma;k}\delta\hat{\boldsymbol{P}}^\dagger_k$. As a result, we obtain

$$\eta_{\omega_{\sigma;k}}\hbar\omega_{\sigma;k}\dot{\hat{N}}_{\sigma;k} = -V\left(\delta\hat{\boldsymbol{J}}_k \boldsymbol{E}^{0*}_{\sigma;k}\hat{q}^\dagger_{\sigma;k} + \delta\hat{\boldsymbol{J}}^\dagger_k \boldsymbol{E}^0_{\sigma;k}\hat{q}_{\sigma;k}\right). \tag{31c}$$

It is evident that with the same work of the field on the "external" current, the sign of change of the photon number depends on the sign of the expression $\boldsymbol{E}^{0*}_{\sigma;k}\left(\partial(\omega^2 \ddot{\varepsilon}_\omega)/\partial\omega^2\right)\boldsymbol{E}^0_{\sigma;k}$.

We emphasize that all the conclusions made in this Section in regard to the notion of the negative energy quantum modes follow from an analysis of the operator equation for the field in the medium and do not contain any phenomenological assumptions.

IX. GENERALIZATION IN THE CASE OF THE CLASSICAL CONTROL FIELD

The above results are based essentially on the assumption that the "nonlinear" corrections arising in the solution of the density operator equations are small. However, these relations have a wider range of applicability. We are talking about the behavior of the quantized field in the presence of a high-power control wave (pump) corresponding to the classical limit. In this case, when the density operator equation is addressed, the pump field operator can be considered an ordinary function of time and coordinates (*c*-number):

$$\hat{\boldsymbol{E}}_d = \boldsymbol{E}_{0d}\, e^{-i(\omega_d t - \boldsymbol{k}_d \boldsymbol{r})} + \boldsymbol{E}_{0d}^*\, e^{i(\omega_d t - \boldsymbol{k}_d \boldsymbol{r})}.$$

In principle, it is possible to obtain a solution of the density operator equation that is linear with respect to the operator of quantized fields $\hat{\boldsymbol{E}}$, but depends in a nonlinear way on the control wave as the parameter (for example, on the intensity $|\boldsymbol{E}_{0d}|^2$). In particular, for the EIT regime, such a solution is radically different from the corresponding linear solution in the absence of the pump (see, e.g., [25, 26, 27, 35]). The polarization induced by quantized field $\hat{\boldsymbol{E}}$ will be determined by expression

$$\hat{\boldsymbol{P}} = \ddot{\chi}(\boldsymbol{E}_{0d})\hat{\boldsymbol{E}},$$

where $\ddot{\chi}(\boldsymbol{E}_{0d})$ is the tensor linear operator which depends on the complex amplitude of the classical pump as the parameter. Formally, ***absolutely all*** of the above results are transferred directly to this case just by virtue of the conservation of the linear connection between the operators of $\hat{\boldsymbol{P}}$ and $\hat{\boldsymbol{E}}$. However, there is one important interpretive difference concerning the physical meaning of expression (22a) for the field energy in the medium. The point is that in this case, the field energy in the medium is not just the sum of the "vacuum" energy of the electromagnetic field and the energy of the medium excitation, but also allows for the reversible energy exchange with the control field. This circumstance was discussed in [53-55] in connection with the EIT effect. It is important to note that the "excess" of free energy is possible, in principle, in the pump-controlled ensemble of atoms. Exactly as in the case of the inverted medium, the negative energy operator can, in principle, be entered. An example of a wave mode that changes the sign of the photon energy in the medium in the presence of the classical pump is a Stokes satellite of the control field that appears during the EIT regime implementation with a relatively high density of atoms (see [42]).

X. CONCLUSIONS

In this paper, we present a self-consistent theory of interaction between the quantum field and an optically dense transparent medium. Examples of the efficiency of the discussed approach are the consistent introduction of the concept of a photon in the medium (using the Bogoliubov transform for the "vacuum" photon creation and annihilation operators) and detailed quantum theory of the negative energy modes. The development of the theory presented in this paper is feasible and useful in the following areas. First, it is the theory of quantum field in inhomogeneous media and the media with spatial dispersion. The latter usually feature the resonant line broadening which can be used in the echoic quantum memory circuits [56-67]. Another interesting and important property of the systems with spatial dispersion is the implementation of quadratic nonlinearity in isotropic media [68].

The next important line of development of the theory is nonlinear analysis that is free of the classical control field approximation. Although the methods of studying multi-photon nonlinear regimes are fairly clear for the media with power-law nonlinearity (see, e.g., [1,2,50,69]), the general case seems to be complicated (see, e.g., the analysis of the EIT regime with quantum pump in the gamma-ray range [70]). However, in many cases it seems sufficient to use the fluctuating classical pump approximation for the analysis of applied problems.

One more interesting range of problems is associated with the development of the methods for allowing the dissipative effects. Straightforward numerical modeling of the relaxation processes (i.e., those which are free of the model approximations) for quantum systems is currently unrealistic. In this respect, the studies offering different sorts of modification of the simplest models seem to be important [41,44-47,71].

ACKNOWLEDGMENTS

This research was funded by the Federal state budgetary institution of science Institute of Applied Physics of the Russian Academy of Sciences. The authors are grateful to A. A. Belyanin, I. D. Tokman, V. V. Kocharovsky, E. V. Radeonychev, and V. A. Mironov for fruitful discussions of various aspects of the theory presented here.

**Appendix A.**

CANONICAL EQUATIONS FOR ELECTROMAGNETIC FIELD IN THE MEDIUM

Here we will construct the model of a continuous medium on the basis of a micro model. To this end, we consider $J$ identical atoms within the volume $V$ at the points with coordinates $\boldsymbol{r} = \boldsymbol{r}_j$, $j = 1,2,\ldots,J$. To the internal degrees of freedom of atoms we assign undisturbed stationary states $|n\rangle$ and energy levels $W_n$. Keeping in mind the medium without spatial dispersion, we assume that the atomic wave functions do not overlap. In this case, one can neglect the requirement of permutation symmetry for a total wave function of the medium (i.e., the function which includes the dependence on the coordinates $\boldsymbol{r}_j$ of the centers of mass) and confine oneself to considering the internal degrees of freedom of atoms (in this regard, see, e.g., [5]). This model exactly corresponds, for example, to the interaction between the radiation and the ensembles of impurity centers and quantum dots [43,72] or groups of valence levels weakly disturbed by the interaction with the neighboring atoms and ions, gases [5, 35, 42, 73], "magnetized" quasi-particles in the Landau level system [37], etc. However, finally, using the above model, we arrive at sufficiently universal equations of quantum electrodynamics of a dielectric medium with frequency dispersion.

We specify the electromagnetic field using the vector and scalar potentials $E(r,t) = -c^{-1}\dot{A} - \nabla\varphi$, $B(r,t) = \nabla \times A$. Using the Coulomb calibration ($\nabla A = 0$), the electromagnetic field is described by the equations

$$\ddot{A} - c^2\nabla^2 A = 4\pi c \dot{P}_\perp, \tag{A.1a}$$

$$\nabla\varphi = 4\pi P_\parallel, \tag{A.1b}$$

where $P_\perp$ and $P_\parallel$ are the vortex and potential components of the polarization vector, respectively:

$$P = P_\perp + P_\parallel, \quad \nabla \times P_\parallel = \nabla P_\perp = 0, \tag{A.2a}$$

$$P = \sum_j \delta(r - r_j) p_j, \tag{A.2b}$$

$$p_j = \sum_{m,n} d_{nm} c_{j;m} c^*_{j;n}. \tag{A.2c}$$

Here, $p_j$ is the dipole moment of the $j$th atom, $c_{j;n}$ is the vector of state of the $j$th atom, $d_{nm}$ is the matrix element of the dipole moment operator, and $d_{nn} = 0$ (i.e., the stationary states of the atoms are not polarized).

Assuming the sizes of the atom to be small compared with the scale of non-uniformity of the electromagnetic field, we specify the matrix element of the Hamiltonian of the $j$th atom within the electric-dipole approximation [1,2,7]:[9]

$$h_{j;nm} = W_n \delta_{nm} - d_{nm} E_j^a, \tag{A.3a}$$

where $E_j^a$ is the field acting on the $j$th atom, i.e., electromagnetic field at the point $r \to r_j$ without the contribution of the $j$th atom [5,6,22,33]. Determine the quantity $E_j^a$ by subtracting from the "total" field $E(r)$ the dipole field $\eta_j(r - r_j)$ of the $j$th atom, which is singular at $r \to r_j$ [22]:

$$E_j^a = \lim_{r \to r_j} (E(r) - \eta_j(r - r_j)), \quad \eta_j = \left(3[(r - r_j) \cdot p_j]\frac{r - r_j}{|r - r_j|^5} - \frac{p_j}{|r - r_j|^3}\right). \tag{A.3b}$$

Formulation of the quantum description of the field usually requires that the original equations of the system are represented in the Hamiltonian form [1, 2, 4, 48]. The Hamiltonian of the "classical field + ensemble of atoms" system and the canonical momenta corresponding to the chosen coordinates can be obtained with the following Lagrangian:

$$L = \int_V \left( \frac{E^2 - B^2}{8\pi} + s - v \right) d^3 r, \quad (A.4a)$$

where

$$s = \sum_{j,n} \delta(\mathbf{r} - \mathbf{r}_j) c^*_{j;n} \left( i\hbar \dot{c}_{j;n} - \sum_m \tilde{h}_{j;nm} c_{j;m} \right), \quad \tilde{h}_{j;nm} = W_n \delta_{nm} - \mathbf{d}_{nm} \mathbf{E},$$

$$v = \frac{1}{2} \sum_j \delta(\mathbf{r} - \mathbf{r}_j) (\boldsymbol{\eta}_j \cdot \mathbf{p}_j) = \frac{1}{2} \sum_j \delta(\mathbf{r} - \mathbf{r}_j) \left( 3 \frac{\left[ (\mathbf{r} - \mathbf{r}_j) \cdot \sum_{n,m} \mathbf{d}_{nm} c_{j;m} c^*_{j;n} \right]^2}{|\mathbf{r} - \mathbf{r}_j|^5} - \frac{\left( \sum_{n,m} \mathbf{d}_{nm} c_{j;m} c^*_{j;n} \right)^2}{|\mathbf{r} - \mathbf{r}_j|^3} \right). \quad (A.4b)$$

---

[9] For simplicity, we do not consider the magnetic effects, i.e., assume that in the CGS system of units the magnetic field $H$ is equal to its induction $B$.

Formally, the divergent terms in the expression for *v*, of course, generate divergences in the equations of the dynamics of the state vector of the *j*th atom as well. However, as in expression (A.3b), the corresponding divergences cancel with the dipole field of the *j*th atom, which is implicitly contained in the "total" field $\mathbf{E}(\mathbf{r})$ and is singular for $\mathbf{r} \to \mathbf{r}_j$.

Lagrangian (A.4a) depends on generalized coordinates $\mathbf{A}(\mathbf{r},t)$, $\varphi(\mathbf{r},t)$, $c_{j;n}(t)$, and $c^*_{j;n}(t)$ and generalized velocities $\dot{\mathbf{A}}(\mathbf{r},t), \dot{c}_{j;n}(t)$, while it does not depend on the velocities $\dot{\varphi}$ and $\dot{c}^*_{j;n}$ (in this regard, see also [4, 74]). The Lagrange equations for the distributed system have the following form:

$$\frac{\partial L}{\partial c^*_{j;n}} = 0, \quad \frac{\partial}{\partial t} \frac{\partial L}{\partial \dot{c}_{j;n}} - \frac{\partial L}{\partial c_{j;n}} = 0, \quad \frac{\delta L}{\delta \varphi} = 0, \quad \frac{\partial}{\partial t} \frac{\delta L}{\delta \dot{\mathbf{A}}} - \frac{\delta L}{\delta \mathbf{A}} = 0, \quad (A.4c)$$

where for the field variables $\mathbf{A}(\mathbf{r},t)$ and $\varphi(\mathbf{r},t)$ the following standard variation derivatives are meant (see [4, 74]):

$$\frac{\delta \int_V \Xi(f_i, \nabla f_i) d^3 r}{\delta f_i(\mathbf{r})} = \frac{\partial \Xi}{\partial f_i} - \nabla \cdot \frac{\partial \Xi}{\partial \nabla f_i}$$

The first of the equations (A.4c) with expressions (A.2b), (A.2c), (A.3b), (A.4a), and (A.4b) taken into account generate the Schrödinger equation corresponding to the Hamiltonian (A.3a) of the *j*th atom in view of relation (A.3b):

$$i\hbar \dot{c}_{j;n} - W_n c_{j;n} = -\mathbf{E}^a_j \sum_m \mathbf{d}_{nm} c_{j;m}; \quad (A.4d)$$

the second of the equations (A.4c) generates a complex-conjugate counterpart of Eq. (A.4d). The latter pair of equations (A.4c) leads to the Poisson equation and the wave equation for the vector potential

$$\nabla^2 \varphi = 4\pi \nabla P \ , \ \ddot{A} + c^2 \nabla \times \nabla \times A = c(4\pi \dot{P} - \nabla \dot{\varphi}),$$

whence Eqs. (A.1a) and (A.1b) follow.

We now pass from the Lagrangian to the Hamiltonian description [74]. Determine the canonical momenta of the atomic excitations $\mu_{j;n}$ and of the electromagnetic field $F$:

$$\mu_{j;n} = \frac{\partial L}{\partial \dot{c}_{j;n}} = i\hbar c^*_{j;n}, \ F = \frac{\delta L}{\delta \dot{A}} = \frac{\dot{A}}{4\pi c^2} + \frac{\nabla \varphi}{4\pi c} - \frac{P}{c}.$$

From the latter relation it follows that within the electric-dipole approximation the field coordinate $A$ corresponds to the canonical momentum $F = -D/4\pi c$, where

$$D = -\frac{\dot{A}}{c} - \nabla \varphi + 4\pi P = E + 4\pi P \tag{A.5a}$$

is the electric displacement vector. Determining the Hamiltonian using the usual relation [4,74] $H = \int_V (F \cdot \dot{A}) d^3 r + \sum_{j,n} \mu_{j;n} \dot{c}_{j;n} - L$, we obtain

$$H(c_{j;n}, \mu_{j;n}, A, D) = \int_V \left( \frac{D^2 + (\nabla \times A)^2}{8\pi} + 2\pi P^2 - DP + \frac{D\nabla \varphi}{4\pi} + w + v \right) d^3 r,$$

where the quantities $P$ and $v$ are determined by Eqs. (A.2b), (A.2c), and (A.4b), in which the replacement $c^*_{j;n} \Rightarrow \mu_{j;n}/i\hbar$ is needed to pass to the canonical pairs of variables $(c_{j;n}, \mu_{j;n})$, and $w$ is the energy density of the atomic system:

$$w = \sum_{j,n} \delta(r - r_j) W_n c_{j;n} c^*_{j;n} = \frac{1}{i\hbar} \sum_{j,n} \delta(r - r_j) W_n c_{j;n} \mu_{j;n}. \tag{A.5b}$$

The Hamiltonian obtained above can be assumed to depend only on the variables $(c_{j;n}, \mu_{j;n}, A, D)$ despite the term $\propto D\nabla\varphi$ in the integrand. Indeed, allowing for the consequence from Eq. (A.5a), namely, $\nabla \varphi = 4\pi P - D - \dot{A}/c = 4\pi P_{//}$, we obtain $DP - D\nabla\varphi/4\pi = DP_\perp$. On the other hand, allowing also for the relation $\nabla D = 0$, it is easy to prove the equality $\int_V D(P - P_\perp) d^3 r = \int_V D\nabla\varphi \, d^3 r / 4\pi = \oint_S (D\varphi) ds / 4\pi$. Since the term reducible to the surface integral does not contribute to the variation of the functional [10] $H$, we obtain the following identity for the variations:

$$\delta \int_V (DP) d^3 r \equiv \delta \int_V (DP_\perp) d^3 r. \tag{A.5c}$$

Thus, the Hamiltonian can be represented in this form or another depending on the convenience for some transformations. Bearing in mind relation (A.5c), we use the following expression:

---

[10] The relation is essentially similar for an ensemble of magnetic dipoles [38].

$$H = \int_V \left( \frac{\boldsymbol{D}^2 + (\nabla \times \boldsymbol{A})^2}{8\pi} + 2\pi \boldsymbol{P}^2 - \boldsymbol{DP} + w + v \right) d^3r . \tag{A.5d}$$

Although the Hamiltonian in form (A.5d) does not depend on the scalar potential $\varphi$, it describes the considered system as a closed one. Indeed, Hamiltonian (A.5d) determines the dynamics of the canonical pairs of variables $(\boldsymbol{A}, \boldsymbol{D})$ and $(c_{j;n}, \mu_{j;n})$, thereby specifying the polarization density $\boldsymbol{P}$. Potential $\varphi$ is excluded in this case. If needed, the potential can be found by any of two equations used above: $\nabla \varphi = 4\pi \boldsymbol{P} - \boldsymbol{D} - \dot{\boldsymbol{A}}/c = 4\pi \boldsymbol{P}_{//}$.

Passing to the quantum description of the field, we consider Hamiltonian (A.5d) as an operator [1,2,4,5,7,48]. To this end, we introduce operators of the quantities $\hat{H}, \hat{\boldsymbol{P}}, \hat{\boldsymbol{A}}, \hat{\boldsymbol{D}}, \hat{w},$ and $\hat{v}$ which enter expression (A.5d):

$$\hat{H} = \int_V \left( \frac{\hat{\boldsymbol{D}}^2 + (\nabla \times \hat{\boldsymbol{A}})^2}{8\pi} + 2\pi \hat{\boldsymbol{P}}^2 - \hat{\boldsymbol{D}}\hat{\boldsymbol{P}} + \hat{w} + \hat{v} \right) d^3r . \tag{A.6a}$$

Operators of the canonically conjugate variables $\hat{\boldsymbol{F}} = -\hat{\boldsymbol{D}}/4\pi c$ and $\hat{\boldsymbol{A}}$ will be specified by a standard commutation relation for the canonically conjugate "momentum – coordinate" pair [1,4,5]:

$$[\hat{F}_\alpha(\boldsymbol{r}',t), \hat{A}_\beta(\boldsymbol{r},t)] = -i\hbar \delta_{\alpha\beta} \delta(\boldsymbol{r} - \boldsymbol{r}'), \tag{A.6b}$$

where $\alpha, \beta = x, y, z$ are the indices of the Cartesian coordinates of the vectors. Determining the operators $\hat{w}, \hat{\boldsymbol{P}}$, and $\hat{v}$, we pass to the secondary quantization representation in expressions (A.2b), (A.2c), (A.4b), and (A.5b), i.e., we replace the dyads $c_{j;m} c^*_{j;n} = c_{j;m} \mu_{j;n}/i\hbar$ by operator matrices $\hat{\rho}_{j;mn}$:

$$\hat{w} = \sum_{j,n} \delta(\boldsymbol{r} - \boldsymbol{r}_j) W_n \hat{\rho}_{j;nn} , \quad \hat{\boldsymbol{P}} = \sum_{j,n,m} \delta(\boldsymbol{r} - \boldsymbol{r}_j) \boldsymbol{d}_{nm} \hat{\rho}_{j;mn} , \tag{A.6c}$$

$$\hat{v} = \frac{1}{2} \sum_j \delta(\boldsymbol{r} - \boldsymbol{r}_j) \left( 3 \frac{[(\boldsymbol{r} - \boldsymbol{r}_j) \cdot \boldsymbol{d}_{nm} \hat{\rho}_{j;mn}]^2}{|\boldsymbol{r} - \boldsymbol{r}_j|^5} - \frac{\left( \sum_{n,m} \boldsymbol{d}_{nm} \hat{\rho}_{j;mn} \right)^2}{|\boldsymbol{r} - \boldsymbol{r}_j|^3} \right) \tag{A.6d}$$

We now consider relations (A.6c) and (A.6d) in more detail. In the secondary quantization representation, the operator of some quantity $f$ has the form $\hat{f} = \sum_{n,m} f_{nm} \hat{\rho}_{mn}$, where $f_{nm}$ is the standard matrix element and $\hat{\rho}_{mn}$ is the operator matrix, which can be specified in different ways. Firstly, one can use the projection operator $\hat{\rho}_{mn} = |n\rangle\langle m|$ [1,5,7]. Another way is to use operators of annihilation and creation of quantum states, $\hat{\rho}_{mn} = \hat{a}_n^\dagger \hat{a}_m$ [1,2,4,39]. Both approaches lead to quite similar results. We will call the operator matrix $\hat{\rho}_{mn}$ a density operator regardless of the presentation method. Note that in some cases by using the formalism of the projection operators one can easily establish specific properties of the density operator. In particular, it is exactly the easiest way to obtain the following commutation relation:

$$[\hat{\rho}_{qp}, \hat{\rho}_{mn}] \equiv [|p\rangle\langle q|, |n\rangle\langle m|] = \left(\delta_{qn}|p\rangle\langle m| - \delta_{mp}|n\rangle\langle q|\right) \equiv \left(\delta_{qn}\hat{\rho}_{mp} - \delta_{mp}\hat{\rho}_{qn}\right). \quad \text{(A.7a)}$$

A commutation relation for the dyads $\hat{a}_n^\dagger \hat{a}_m$, which corresponds to Eq. (A.7a) was obtained in [37,41]:

$$[\hat{\rho}_{qp}, \hat{\rho}_{mn}] \equiv [\hat{a}_p^\dagger \hat{a}_q, \hat{a}_n^\dagger \hat{a}_m] = \left(\delta_{qn}\hat{a}_p^\dagger \hat{a}_m - \delta_{mp}\hat{a}_n^\dagger \hat{a}_q\right) \equiv \left(\delta_{qn}\hat{\rho}_{mp} - \delta_{mp}\hat{\rho}_{qn}\right); \quad \text{(A.7b)}$$

it was shown that this result does not depend on whether the operators $\hat{a}_m$ and $\hat{a}_n^+$ satisfy the boson commutation relations when $[\hat{a}_m, \hat{a}_n] = [\hat{a}_m^\dagger, \hat{a}_n^\dagger] = 0$, $[\hat{a}_m, \hat{a}_n^\dagger] = \delta_{mn}$ or fermion relations for which $[\hat{a}_m, \hat{a}_n]_+ = [\hat{a}_m^\dagger, \hat{a}_n^\dagger]_+ = 0$, $[\hat{a}_m, \hat{a}_n^\dagger]_+ = \delta_{mn}$ (here, $[\hat{u}, \hat{v}]_+ = \hat{u}\hat{v} + \hat{v}\hat{u}$).

The following generalization of Eqs. (A.7a) and (A.7b) is valid for an ensemble of atoms [37]:

$$[\hat{\rho}_{j;qp}, \hat{\rho}_{i;mn}] = \delta_{ji}\left(\delta_{qn}\hat{\rho}_{j;mp} - \delta_{mp}\hat{\rho}_{j;qn}\right), \quad \text{(A.7c)}$$

where the density operator $\hat{\rho}_{j;mn}$ for the $j$th atom was introduced. The presence of the symbol $\delta_{ji}$ in Eq. (A.7c) reflects the fact that the operator $\hat{\rho}_{j;mp}$ was defined in the space of states of "its own" atom and has no effect on the states of other atoms. Obviously, the latter circumstance is in no way related to the presence or absence of interaction between atoms (including interaction through a collective field). Similarly, the atomic operators commute with the field operators.

Consider the Heisenberg equation for the density operator:

$$\dot{\hat{\rho}}_{j;mn} = \frac{i}{\hbar}[\hat{H}, \hat{\rho}_{j;mn}]. \quad \text{(A.8a)}$$

Using expressions (A.6a), (A.6c), (A.6d), and (A.7c), we arrive at the following evolution equation:

$$\dot{\hat{\rho}}_{j;mv} = -\frac{i}{\hbar}\sum_{v}\left(\hat{h}_{j;mv}\hat{\rho}_{j;vn} - \hat{\rho}_{j;mv}\hat{h}_{j;vn}\right), \quad \text{(A.8b)}$$

where

$$\hat{h}_{j;mv} = W_m\delta_{mv} - \hat{E}_j^a \mathbf{d}_{mv}, \quad \text{(A.8c)}$$

$\hat{E}_j^a(\mathbf{r}, t)$ is the operator of the field acting on the $j$th atom [5,6,33]:

$$\mathbf{E}_j^a = \lim_{\mathbf{r}\to\mathbf{r}_j}\left(\hat{\mathbf{D}} - 4\pi\hat{\mathbf{P}} - \hat{\boldsymbol{\eta}}_j(\mathbf{r} - \mathbf{r}_j)\right), \quad \hat{\boldsymbol{\eta}}_j = \left(3\left[(\mathbf{r} - \mathbf{r}_j)\cdot \sum_{n,m}\mathbf{d}_{nm}\hat{\rho}_{j;mn}\right]\frac{\mathbf{r} - \mathbf{r}_j}{|\mathbf{r} - \mathbf{r}_j|^5} - \frac{\sum_{n,m}\mathbf{d}_{nm}\hat{\rho}_{j;mn}}{|\mathbf{r} - \mathbf{r}_j|^3}\right). \quad \text{(A.8d)}$$

The equation for atomic density operator (A.8b) has the form of a standard von Neuman equation, although the initial equation of motion for Heisenberg operator (A.8a) has a different sign before the corresponding term with commutator. This fact was mentioned in [37, 41]. (To avoid a misunderstanding, we recall once again that unlike the standard density matrix, whose elements are complex and real numbers, the matrix elements of the atomic density operator $\hat{\rho}_{j;mn}$ introduced above are operators). We also recall that commutation relations for the time-dependent Heisenberg operators always remain the same as for the Schrödinger operators. The corresponding proof (which is rather evident, however) is given in, e.g., [4].

To pass to the model of continuous medium, it is needed to "get rid" of the space coordinates $r_j$ of individual atoms. For this purpose, we use averaging over an ensemble of positions of atoms in space. Formulating the averaging procedure, we introduce the multi-particle probability density of positions of atoms in space $\zeta(r_1,...r_J)$, which satisfies the normalization $\int_V d^3r_1...\int_V d^3r_J \zeta(r_1,...r_J) = 1$. If we are speaking of atoms fixed at the lattice sites, then all the vectors $\Delta r_{j \neq j_0} = r_j - r_{j_0}$ are rigid. In this case, the continuum approximation corresponds to the assumption about an ensemble of realizations for the position of some "reference site" $r_{j_0}$ within the volume $V$. The requirement of permutation symmetry for the total wave function of the medium can affect, in general, the form of the distribution $\zeta(r_1,...r_J)$.[11] However, the further consideration does not depend on this circumstance.

Consider some quantity $\Theta(r,r_1,...r_J)$, which depends on the relative positions of atoms.[12] Ensemble average of possible distributions of atoms in space $\overline{\Theta}$ is defined by the relation

$$\overline{\Theta}(r) = \overline{\Theta(r,r_1,...r_J)} \equiv \int_V d^3r_1...\int_V d^3r_J \Theta(r,r_1,...r_J) \zeta(r_1,...r_J).$$

We will use this averaging procedure for Hamiltonian (A.6a) and relations (A.6b), (A.6c), and (A.7c). Passing to the model of a transparent continuous medium, the terms that are quadratic with respect to fluctuations $\delta\Theta = \Theta - \overline{\Theta}$ should be neglected during averaging since the relative magnitude of respective fluctuations decreases with increasing number of particles in the averaging volume [76]: $\overline{(\delta\Theta)^2}/\overline{\Theta}^2 \propto J^{-1}$. By so doing, we lose, of course, a variety of effects. These include light scattering by the fluctuating parameters of the medium or Bragg scattering by lattice,[13] inhomogeneous broadening of resonant lines due to interaction of atoms, etc. Effects of this kind can be taken into account (if necessary) phenomenologically, by adding additional terms in the overall dy-namics equations of the field and the medium. A similar phenomenological approach is often used to allow for dissipative effects in open systems (see, e.g., [24,25,42,37,40,41,42]).

---

[11] For example, the density distribution of ultracold gas in a trap is different for bosons and fermions [75].
[12] Exactly these quantities are, in particular, the electromagnetic fields at point $r$.
[13] Naturally, the scattering effects put a lower limit on the interval of the wavelengths for which the approximations of a continuous transparent medium is valid.

A fundamental point in the transition to the limit of a continuous medium is the fact that the difference between the local field $\hat{\boldsymbol{E}}_j^a$ acting on a given atom and the average macroscopic field $\overline{\boldsymbol{E}} = \overline{\boldsymbol{D}} - 4\pi\overline{\hat{\boldsymbol{P}}}$ is, in general, a random variable with a nonzero average value [5,6,22,33]. Generally, connection between the local and macroscopic fields in the continuum approximation is given by $\overline{\hat{\boldsymbol{E}}^a} = \overline{\boldsymbol{E}} + \alpha\overline{\hat{\boldsymbol{P}}}$, where the constant $\alpha$ depends on relative positions of the elementary dipoles, i.e., on the form of the distribution $\zeta(\boldsymbol{r}_1,...\boldsymbol{r}_J)$. The most widespread Lorenz-Lorentz model corresponds to the value $\alpha = 4\pi/3$. Applicability of this model within the framework of a quantum-electrodynamic description was demonstrated in [33] for the two-level system. In terms of the Hamilton approach, the local-field effect can be taken into account by selecting in the Hamiltonian of system (A.6a) the terms that allow for the energy of the "near" dipole-dipole interaction of atoms and the energy of interaction between the dipoles and the field of the distant atoms and external field sources.[14] A similar procedure was described in sufficient detail for the ensemble of magnetic dipoles [38]. In what follows we give the final form of the Hamiltonian in the continuum approximation for the system considered. Since we will use only averaged operators and quantities in the further analysis, we will omit a special symbol (overbar) to denote them:

$$\overline{\hat{H}},\overline{\hat{\boldsymbol{A}}},\overline{\hat{\boldsymbol{D}}},\overline{\hat{w}},\overline{\hat{\boldsymbol{P}}},\overline{\hat{\rho}}_{mn} \Rightarrow \hat{H},\hat{\boldsymbol{A}},\hat{\boldsymbol{D}},\hat{w},\hat{\boldsymbol{P}},\hat{\rho}_{mn};$$

$$\hat{H} = \int_V \left( \frac{\hat{\boldsymbol{D}}^2 + (\nabla \times \hat{\boldsymbol{A}})^2}{8\pi} + \hat{w} - \hat{\boldsymbol{D}}\hat{\boldsymbol{P}} + \left(2\pi - \frac{\alpha}{2}\right)\hat{\boldsymbol{P}}^2 \right) d^3r \quad (A.9a)$$

$$\hat{w}(\boldsymbol{r},t) = \sum_n W_n \hat{\rho}_{nn}(\boldsymbol{r},t), \quad \hat{\boldsymbol{P}}(\boldsymbol{r},t) = \sum_{n,m} \boldsymbol{d}_{nm}\hat{\rho}_{mn}(\boldsymbol{r},t), \quad (A.9b)$$

$$[\hat{\rho}_{qp}(\boldsymbol{r},t),\hat{\rho}_{mn}(\boldsymbol{r}',t)] = \delta(\boldsymbol{r}-\boldsymbol{r}')(\delta_{qn}\hat{\rho}_{mp}(\boldsymbol{r},t) - \delta_{mp}\hat{\rho}_{qn}(\boldsymbol{r},t)), \quad (A.9c)$$

where

$$\hat{\rho}_{mn}(\boldsymbol{r},t) = \overline{\sum_j \delta(\boldsymbol{r}-\boldsymbol{r}_j)\hat{\rho}_{j;mn}(t)}. \quad (A.9d)$$

The following operator equality corresponds to the relation (A.5c):

$$\left[\int_V (\hat{\boldsymbol{D}}\hat{\boldsymbol{P}})d^3r, \hat{\boldsymbol{A}}\right] = \left[\int_V (\hat{\boldsymbol{D}}\hat{\boldsymbol{P}}_\perp)d^3r, \hat{\boldsymbol{A}}\right]. \quad (A.10)$$

The identity (A.10a) can be proved, in particular, by expanding the vector fields $\hat{\boldsymbol{D}}$ and $\hat{\boldsymbol{P}}$ over plane waves and making use of the condition $\nabla\hat{\boldsymbol{D}} = 0$ (see Section **II**).

---

[14] That is, separated by a distance which is large compared with the typical distance between the elementary dipoles [5,6,22,33,38].

Using the identities $\hat{\boldsymbol{D}} = \hat{\boldsymbol{E}} + 4\pi\hat{\boldsymbol{P}}$ and $\hat{\boldsymbol{B}} = \nabla \times \hat{\boldsymbol{A}}$, it is easy to obtain from Eq. (A.9a) another, physically clear form for energy operator (2a) presented in Section **II**.

Commutation relation (A.9c) is the generalization of expression (A.7c) to the case of a spatially distributed system. This commutation relation was also obtained in [37] by averaging over a physically small volume. Commutation relation (A.6b) remains valid for the field operators $\hat{\boldsymbol{A}}, \hat{\boldsymbol{D}}$ after the averaging.

**Appendix B.**

SOME FEATURES OF THE OPERATOR EQUATIONS

Heisenberg operators act on the constant (initial) quantum state $|\Psi\rangle$ of the system. For simplicity, we assume that at the initial time one have the state $|\Psi\rangle = |\Psi_F\rangle|\Psi_A\rangle$.

In quantum theory of the field, initial values of the field operators are often chosen for convenience in such a way that the initial quantum state $|\Psi_F\rangle$ is the vacuum state $|0_F\rangle$ [4,48]. Transition from the "real" initial state to $|0_F\rangle$ is possible due to simultaneous changes in the initial state vector and the initial values of the operators using appropriate unitary transformations. The same applies to the choice of the initial values of the medium state vector $|\Psi_A\rangle$ and the density operator. To see this, we determine the initial value of the operator $\hat{\rho}_{mn}(\boldsymbol{r},t)$ using relation (A.9d):

$$\hat{\rho}^0_{mn}(\boldsymbol{r}) = \overline{\sum_j \delta(\boldsymbol{r}-\boldsymbol{r}_j)\hat{\rho}^S_{j;mn}}, \tag{B.1}$$

where $\hat{\rho}^S_{j;mn}$ is the Schrödinger (constant) operator. Then we pass, for example, from the "real" initial state $|\Psi_A\rangle = \prod_j |\Psi_j\rangle$ of the ensemble of non-interacting atoms to the initial state $|0_A\rangle = \prod_j |0_j\rangle$, in which all the atoms are in the ground states. One should pass from Eq. (B.1) to the operator

$$\hat{\tilde{\rho}}^0_{mn}(\boldsymbol{r}) = \overline{\sum_j \delta(\boldsymbol{r}-\boldsymbol{r}_j)\hat{U}^\dagger_j\hat{\rho}^S_{j;mn}\hat{U}_j}, \tag{B.2}$$

where $\hat{U}_j$ is the unitary operator determined by the relation $|\Psi_j\rangle = \hat{U}_j|0_j\rangle$. Obviously, the commutation and correlation relations for the operators $\hat{\rho}^0_{mn}$ and $\hat{\tilde{\rho}}^0_{mn}$ coincide by virtue of the properties of the unitary operators $\hat{U}^\dagger_j\hat{U}_j = 1$. Thus, the choice of the initial state vector $|\Psi_A\rangle$ and respective initial density operator has, in a sense, symbolic value and is determined by the ease of description. Direct physical meaning is in the corresponding mean values at the initial time. Therefore, it is convenient to specify the initial density operator $\hat{\rho}^0_{mm}(\boldsymbol{r})$ not directly, but via mean values at the initial time that are necessary for the problem. In particular, we give the result which is important for the further discussion, namely, that if all atoms are in stationary states at the initial time, then the relation

$$\langle \Psi_A | \hat{\rho}^0_{mm}(\boldsymbol{r}) \hat{G} | \Psi_A \rangle = N_m(\boldsymbol{r}) \langle \Psi_A | \hat{G} | \Psi_A \rangle \tag{B.3}$$

is valid for any operator $\hat{G}$. To prove the property (B.3) one can use the representation of the operator $\hat{\rho}^0_{mm}$ in the form (B.1), where $\hat{\rho}^S_{j;mm} = |m_j\rangle\langle m_j|$ should be substituted.